\documentclass[preprints,review,accept,moreauthors]{Definitions/mdpi} 

\usepackage{amsmath}
\usepackage{amssymb}
\usepackage{graphicx,bbm,mathrsfs}
\usepackage{nicefrac}
\usepackage{slashed}
\usepackage{bbm}
\usepackage{geometry}
\usepackage{empheq}
\usepackage{stackrel}
\usepackage{ulem}
\usepackage{setspace}
\usepackage{relsize}
\usepackage{empheq}
\usepackage{pifont}% http://ctan.org/pkg/pifont
\usepackage{mathtools}
\usepackage{enumitem}
\usepackage{dcolumn}   % needed for some tables
\usepackage{bm}        % for math
\usepackage{graphicx,mathrsfs}
\usepackage{nicefrac}
\usepackage{multirow}
\usepackage{color}
\usepackage{mathtools}
\usepackage{makecell}
\usepackage{environ} 
\usepackage{lipsum} 
\usepackage{wasysym}
\usepackage{hhline,colortbl}

\usepackage{tikz}
\usepackage{graphicx}
\usepackage{tensor}
\usepackage{dsfont}
\usepackage{simpler-wick}
\usepackage{accents}

\newcommand*{\dt}[1]{%
  \accentset{\mbox{\bfseries .}}{#1}}

\newcommand{\tr}{\operatorname{tr}}

\usepackage{booktabs}
\usepackage{mdframed}
\usepackage{xcolor}
\definecolor{EqFrame}{RGB}{235,245 ,250 }

\newcommand{\be}{\begin{equation}}
\newcommand{\ee}{\end{equation}}
\newcommand{\bea}{\begin{eqnarray}}
\newcommand{\eea}{\end{eqnarray}}
\newcommand{\SM}{{\textrm{SM}}}

\firstpage{1} 
\pubvolume{1}
\issuenum{1}
\articlenumber{0}
\pubyear{2026}
\copyrightyear{2026}
\datereceived{ } 
\daterevised{ } % Comment out if no revised date
\dateaccepted{ } 
\datepublished{ } 
\Title{Linear Dilaton in Cosmology and Particle Physics}

\Author{Eugenio Meg\'{\i}as $^{1}$\orcidA{}, and Mariano Quir\'os $^{2}$\orcidB{}*}

\AuthorNames{Eugenio Meg\'{\i}as, and Mariano Quir\'os}

\address{%
$^{1}$ \quad Departamento de F\'{\i}sica At\'omica, Molecular y Nuclear and Instituto Carlos I de F\'{\i}sica Te\'orica y Computacional, Universidad de Granada, Avenida de Fuente Nueva s/n, 18071 Granada, Spain; emegias@ugr.es\\
$^{2}$ \quad Institut de F\'{\i}sica d'Altes Energies (IFAE) and The Barcelona Institute of  Science and Technology (BIST), Campus UAB, 08193 Bellaterra, Barcelona, Spain; quiros@ifae.es}

\corres{Correspondence: quiros@ifae.es}

\abstract{A warped extra dimension in a five-dimensional (5D) anti-de Sitter (AdS) background was introduced in 1999 by Lisa Randall and Raman Sundrum to solve the gauge hierarchy problem in particle physics. As a bonus, a holographic interpretation in terms of four-dimensional (4D) conformal field theories (CFTs) was found. Interestingly enough, another 5D background, linear dilaton (LD), was found to have a holographic interpretation in terms of Little String Theory. In this review, we will show how a set of 5D backgrounds, parametrized in terms of a real parameter $\nu$, generalizes both theories and gives rise, in particular, to AdS for $\nu=0$ and to LD for $\nu=1$. Furthermore, working in the 5D theory, we will consider applications of the LD background to: (i) particle physics, so that the 5D Planck scale can be lowered to sub-Planckian values, and (ii) Brane World Cosmology (BWC), based on the appearance of an extra vacuum characterized by a 5D black hole. In all cases, we find a gapped continuum for bulk propagating fields, which makes a connection with unparticles. In the case of BWC, we also point out the existence of a pressureless holographic fluid that could play the role of dark matter (DM), with feeble (gravitational) interactions with the Standard Model (SM), decoupled from the thermal SM bath, and generated by a freeze-in mechanism after inflation. We also point out the additional possibility of identifying DM with a long-lived, feebly interacting massive graviton, as an isolated resonance generated by radiative corrections to the continuum graviton propagator~self-energy. }

\begin{document}

%%%%%%%%%%%%%%%%%%%%%%%%%%%%%%%%%%%%%%%%%%
%\setcounter{section}{-1} %% Remove this when starting to work on the template.
\section{Introduction}

Theories with a warped extra dimension were proposed by Lisa Randall and Raman Sundrum~\cite{Randall:1999ee} as a solution to the hierarchy problem. These theories contain an anti-de Sitter (AdS) background metric and two branes: an ultra-violet (UV) brane and an infra-red (IR) brane where the Higgs is either located, or at least toward which its profile is localized if the Higgs is propagating in the bulk. Moreover, the theory contains a massless radion, while the brane distance is undefined, as the radion potential is flat. To make the radion massive, the conformal invariance of the theory needs to be explicitly broken by a bulk potential $V(\phi)$ and brane potentials $\lambda_a(\phi)$, in terms of a stabilizing bulk field $\phi$, which then gives a mass to the radion and introduces an extra excitation around the interbrane separation~\cite{Goldberger:1999uk}. Moreover, the same authors proposed a theory with an infinite extra warped dimension and a single brane where the Standard Model (SM) sector is localized~\cite{Randall:1999vf}. This model does not solve the hierarchy problem (one would require, e.g.,~supersymmetry in the brane to achieve this task) but gave rise to braneworld cosmologies, where the location of the brane is subject to the Friedmann equations and plays the role of the scale factor of the universe. An interesting property of these theories, with AdS geometry, is that they have an equivalent dual defined in the boundary of the AdS space, the holographic dual, known to be a non-perturbative conformal field theory (CFT)~\cite{Aharony:1999ti}.

Interestingly enough, there is another warped geometry for which the holographic dual is known: the linear dilaton (LD) background metric, for which the holographic dual is Little String Theory (LST), a six-dimensional theory compactified on a torus~\cite{Aharony:1998ub,Aharony:1999ks}. The particle physics phenomenology of LD metrics was carried out in Refs.~\cite{Antoniadis:2001sw,Antoniadis:2011qw}, mainly with the goal of lowering the string scale to the TeV to find experimental signatures at present and future colliders. In the case considered in Refs.~\cite{Antoniadis:2001sw,Antoniadis:2011qw}, the warped geometry does not solve the hierarchy problem between the TeV and the Planck scales, which is left to be solved by string theory.
In this review, we will show that there are other possibilities for the LD metric in particle physics, for which the string scale is some orders of magnitude below the Planck scale, while the low-energy scale is at the TeV, say midway between the original LD proposal of Refs.~\cite{Antoniadis:2001sw,Antoniadis:2011qw} and the AdS theory~\cite{Randall:1999ee}. Moreover, we have explored braneworld cosmologies based on the LD background, which provides an interesting possibility for the holographic fluid to describe cosmological dark matter.

The contents and summary of this review are as follows. In Section~\ref{sec:dilaton-gravity}, we will present the general structure of the dilaton-gravity model in the presence of a general class of metrics which contain, as particular cases, the AdS and LD versions. The background vacua in the presence and absence of a 5D black hole (BH) are also computed. Applications to particle physics are presented in Section~\ref{sec:particlephysics}, where we show that there is a solution for which $M_5\sim 10^{-4}M_4$, where $M_5$ and $M_4$ are the Planck scales in five and four dimensions, respectively. The warp factor solves the hierarchy between the low-energy scale $\eta\sim$, TeV scale and $M_5$, while string theory has to be responsible for fixing the hierarchy between $M_5$ and $M_4$. It is also proven that there is a singularity in proper coordinates at $y=y_s=1/\eta$, and, if the interval is extended up to the singularity $(y \in [0,y_s])$,
the bulk propagator is a gapped continuum for $\sqrt{s}>m_g=3\eta/2$. Applications to braneworld cosmology are carried out in Section~\ref{sec:braneworld} for the general class of models presented in Section~\ref{sec:dilaton-gravity}. In the presence of the BH, there is an unconventional cosmology with a holographic fluid which red-shifts with the universe expansion in different ways. In particular, for the case of AdS, the holographic fluid is dark radiation, while for LD, it is pressureless matter. The properties of the holographic fluid for the case of the LD background are studied in Section~\ref{sec:holo_fluid_DM}, where it is shown that it can be a constituent (or the whole) of the cosmological dark matter (DM) of the universe. As the holographic fluid is feebly coupled to matter, under the assumption that the cosmological inflaton is only coupled to the SM, the fluid density will be zero after inflation and at the reheating temperature. We prove that, by the leakage of gravitons into the bulk, its density is restored by the freeze-in mechanism; it is UV-dominated, and then it is controlled by the reheating temperature $T_R$ after inflation. Another possibility provided by the same class of models is explored in Section~\ref{sec:isolated}. The spectrum of the bulk graviton is a gapped continuum, similar to the case of gapped unparticles. We learned from the case of general unparticle propagators that interaction with ordinary matter can provide self-energies that, when resummed, can give rise to an isolated pole in the second Riemann sheet and a pole mass near (and below) the mass gap. In the case of the continuum of gravitons, they behave as unparticles with dimension $d_\mathcal U=3/2$ and couple to the SM fields through their energy-momentum tensors. We have proved that, indeed, an isolated feebly interacting long-lived graviton can appear in the spectrum. We have studied the conditions, from all theoretical and cosmological constraints, under which this isolated state can be cold dark matter. Interestingly enough, it is required that a heavy field, beyond the SM, be present with a mass of $\sim$$M_5$. We have proposed such a field as the cosmological inflaton localized in the brane. The final result is that the isolated graviton can be DM, provided that its mass is below the MeV range. Moreover, as an existence proof, we have presented a simple inflationary model localized in the brane, which satisfies all required properties and has correct predictions of cosmological observables. Interestingly enough, the model evades Lyth's bound about super-Planckian excursions. A discussion of the results and future perspectives is carried out in Section~\ref{sec:discussion}. Finally, for completeness, in the appendices, we compare the orbifold and interval pictures, Appendix~\ref{sec:interval-orbifold}, and present results for a general class of $\nu$-models which are asymptotically AdS, Appendix~\ref{sec:AAdS}.

\section{A Class of Dilaton-Gravity Models}
\label{sec:dilaton-gravity}

We will here consider a slice of 5D spacetime between a brane at the value $y_0=0$ in proper coordinates $y$, the UV boundary brane, and a (possible) admissible singularity~\cite{Gubser:2000nd} placed at $y_s$, a value which has to be determined dynamically. In addition, we will eventually introduce an IR brane at $y_1<y_s$, responsible for electroweak breaking, where we will assume the SM sector to be localized. The considered interval in the downstairs picture $[0,y_s]$ is completely equivalent to the orbifold $S^1/\mathbb{Z}_2$ in the upstairs picture. In the absence of brane terms in the action, even (odd) fields in the orbifold picture have Neumann (Dirichlet) boundary conditions in the interval picture.  In this article, we will be working in the interval picture. The relation between interval and orbifold pictures is described in Appendix~\ref{sec:interval-orbifold}.

The 5D action $\mathcal S$ of the model, including the stabilizing bulk scalar $\phi$, with mass dimension $3/2$, reads as
\begin{eqnarray}
\mathcal S &=& \int d^4x\int_0^{y_s}dy \sqrt{|\det g_{MN}|} \left[ -\frac{1}{2\kappa^2} R + \frac{1}{2} g^{MN}(\partial_M \phi)(\partial_N \phi) - V(\phi) \right]\nonumber \\
&-& \sum_{a} \int_{B_a} d^4x \sqrt{|\det \bar g_{\mu\nu}|} \lambda_a(\phi)  
 -\frac{1}{\kappa^2} %\sum_{\begin{equation}ta=0,s} 
 \int_{B_0} d^4x \sqrt{|\det \bar g_{\mu\nu}|} K_0  \,, \label{eq:action}
\end{eqnarray}
where $\kappa^2\equiv1/M^3_5$, $M_5$ is the 5D Planck scale, $V(\phi)$ and $\lambda_a(\phi)$ are the bulk and brane potentials of the scalar field $\phi$, and the index is $a=0$, with 1 referring to the UV and IR branes, respectively. The IR brane is responsible for the generation of the IR scale $\sim$ TeV and contains the brane Higgs potential, which spontaneously breaks the electroweak symmetry, thus solving the hierarchy problem between $M_5$ and the TeV scale for the considered value of $A(y_1)$, as we will see. 
In Equation~(\ref{eq:action}), the 4D-induced metric is $ \bar g_{\mu\nu}=e^{-2A(y)}\eta_{\mu\nu}$, where the Minkowski metric is given by $\eta_{\mu\nu} =\textrm{diag}(1,-1,-1,-1)$. 
Furthermore, the last term in Equation~(\ref{eq:action}) corresponds to the standard Gibbons--Hawking--York boundary action~\cite{York:1972sj,Gibbons:1976ue}, where $K_{0}$ is the extrinsic UV curvature. In terms of the metric of Equation~(\ref{eq:metric}), the extrinsic curvature term simplifies to~\cite{Megias:2018sxv} $K_{0} =  -4 A^\prime(y_{0})$. It is worth noting that the determinant's contribution inherently regulates and cancels any extrinsic curvature effects at the singularity.

\subsection{No Black Hole Background}

We will now consider a 5D metric, which generalizes the AdS metric and is given by
\begin{equation}
ds^2 =g_{MN}dx^M dx^N\equiv e^{-2A(y)} \eta_{\mu\nu} dx^\mu dx^\nu - dy^2 \,.  \label{eq:metric}  
\end{equation}
The equations of motion (EoM) read then as
\begin{eqnarray}
&&A^{\prime\prime}
= \frac{\kappa^2}{3} \phi^{\prime \, 2} + \frac{\kappa^2}{3} \sum_a \lambda_a(\phi) \delta(y - y_a)  \,, \label{eq:eom1}\\
&&A^{\prime\, 2} 
= -\frac{\kappa^2}{6} V(\phi) + \frac{\kappa^2}{12} \phi^{\prime\, 2} \,,  \label{eq:eom2}\\
&&\phi^{\prime\prime} = 4 A^\prime \phi^\prime +  V^\prime(\phi) + \sum_\alpha \lambda_a^\prime(\phi) \delta(y - y_a)  \,, \label{eq:eom3}
\end{eqnarray}
where the prime symbol $(\,{}^\prime\,)$ will hereafter stand for the derivative of a function with respect to its argument.
The EoM in the bulk can also be written in terms of the superpotential $W(\phi)$ as~\cite{DeWolfe:1999cp}
\begin{equation}
\phi^\prime = \frac{1}{2} \frac{\partial W}{\partial \phi} \,, \qquad A^\prime = \frac{\kappa^2}{6} W \,, \label{eq:phiA}
\end{equation}
and
\begin{equation}
V(\phi) = \frac{1}{8} \left( \frac{\partial W}{\partial \phi} \right)^2 - \frac{\kappa^2}{6} W^2(\phi) \,. \label{eq:V}
\end{equation}

It can be easily checked that if the superpotential $W$ grows faster than $\phi^2$ at large $\phi$, the profile $\phi(y)$ diverges at a finite value $y \equiv y_s$~\cite{Cabrer:2009we}. Moreover, the Ricci scalar can be written~as
\begin{equation}
  R = %20 A^{\prime \, 2} - 8 A^{\prime\prime}
  \frac{5}{9} \kappa^4 W^2(\phi) - \frac{2}{3} \kappa^2 \left( \frac{\partial W}{\partial \phi}  \right)^2  \,,
\end{equation}
so that the curvature in general diverges at $y = y_s$. The physical interpretation is that spacetime ends at $y_s$.
  
The brane potential terms in the EoM are responsible for boundary and jumping conditions for the fields in the branes. In particular, by integrating the equations in a neighborhood of each brane and using Equation~(\ref{eq:phiA}), we get on the UV boundary
\begin{equation}
W(\phi(y_0)) = \lambda_0(\phi(y_0)) \,,\qquad W^\prime(\phi(y_0)) = \lambda_0^\prime(\phi(y_0)) \,, \label{eq:BCUV}
\end{equation}
where the $\mathbb Z_2$ orbifold conditions have been used. On the other hand, in case there is an IR brane at $y = y_1$, we have to impose continuity conditions for $W(\phi)$ and $W^\prime(\phi)$, i.e.,~
\begin{equation}
\Delta W(\phi(y_1))=0 \,, \qquad \Delta W^\prime(\phi(y_1))=0 \,, 
\label{eq:BCIR}
\end{equation}
where $\Delta X$ is the jump of the function $X$ when crossing the brane. Simple brane potentials satisfy the boundary, Equation~(\ref{eq:BCUV}), and jumping, Equation~(\ref{eq:BCIR}), conditions, and fixing the values of $\phi$ dynamically at the branes, i.e.,~$v_a\equiv \phi(y_a)$, is given by
\begin{equation}
\lambda_0(\phi)=W(\phi)+\frac{1}{2}\gamma_0(\phi-v_0)^2\,, \qquad  \lambda_1(\phi)=\frac{1}{2}\gamma_1(\phi-v_1)^2  \,.
\label{eq:lambda01}
\end{equation}

By performing an integration by parts in the bulk of the 5D gravitational Lagrangian, and subsequently enforcing the background EoM, boundary-localized potential terms emerge in the form of $\mp e^{-4A(y_a)}W$~\cite{Gubser:1999vj}, where the $\mp$ sign corresponds to the boundary $y_a=(y_0,\,y_s)$. While this contribution vanishes at the singularity at $y=y_s$, it yields a non-trivial piece at the UV boundary, thereby modifying the effective UV brane potential~into
\begin{equation}
U_0(\phi)=\lambda_0(\phi)-W(\phi)=\frac{1}{2}\gamma_0(\phi-v_0)^2  \,, \label{eq:U0}
\end{equation}
which is dynamically minimized for $\phi=v_0$. Moreover, on the IR brane, there is no such boundary contribution, and there, the effective potential is $U_1(\phi)=\lambda_1(\phi)$, which is minimized for the value of $\phi=v_1$.

For convenience, we will define the dimensionless field $\bar\phi\equiv\kappa\phi/\sqrt{3}$. The properties of the 5D theory depend on the superpotential behavior in the limit $\bar\phi\to\infty$~\cite{Cabrer:2009we}. %In particular when the asymptotic superpotential behavior is exponential $e^{\nu\bar\phi}$, for $\nu<1$ the spectrum is continuous without mass gap, for $\nu>1$ there is a mass gap and a discrete spectrum, and for the critical value $\nu=1$ the spectrum is continuous with a mass gap.
In particular, when the asymptotic superpotential behavior is exponential $e^{\nu\bar\phi}$, there are the following possible situations: (i)~for $\nu<1$, the spectrum is continuous without a mass gap,  (ii) for $\nu>1$, there is a mass gap and a discrete spectrum, and (iii) for the critical value $\nu=1$, the spectrum is continuous with a mass gap. These cases can be covered by the superpotential and potential given by 
\begin{equation}
W(\bar\phi) = \frac{6k}{\kappa^2}e^{\nu\bar\phi},\qquad V(\bar\phi) = -\frac{3k^2}{2\kappa^2}(4-\nu^2)e^{2\nu\bar\phi} \,.
\label{eq:W_lineardilaton_nu}
\end{equation}
We can see that the case $\nu=0$ provides a constant bulk potential and corresponds to the AdS background. The case $\nu=1$ is the so-called linear dilaton background, while the limiting case $\nu=2$ and the case $\nu > 2$ correspond to a zero and positive bulk potential, respectively. The two cases, $\nu = 2$ and $\nu > 2$, are excluded, as they lead to a bad singularity according to Gubser's criterion~\cite{Gubser:2000nd}. Potentials with exponential behavior in the fields, as in Equation~(\ref{eq:W_lineardilaton_nu}), are ubiquitous in supergravity, and they are quite common in, for instance, consistent truncations from 10-dimensional supergravity theories: the simplest example would be to reduce the string theory on a sphere and keep the dilaton, thus leading, in general, to a single exponential. There are also other stringy origins of this potential; see, e.g.,~Refs.~\cite{Cvetic:2000pn,Dong:2012se,Perlmutter:2012he} for the relevant literature on the subject. An analysis and classification of the top-down deductions of the superpotential presented above is intended to be presented in a future work.~\endnote{We would like to thank Ant\'on F. Faedo, Sylvain Fichet, Carlos Hoyos and Jes\'us Huertas for helpful discussions and clarifications on this issue.}

  %We are currently studying the top-down deduction of the superpotential presented above, and the analysis is intended to be presented in a future work.}

The solution of the EoM (\ref{eq:phiA}) provides the background in proper coordinates, given by
\begin{align}
A(y)&=-\frac{1}{\nu^2}\log\left(1-\frac{y}{y_s} \right) \,, \label{eq:sol_A_T0} \\
\bar\phi(y)&=-\frac{1}{\nu}\log[\nu^2 k(y_s-y)] \,, \label{eq:sol_T0}
\end{align}
where we conventionally fix the metric at $y=0$ as $A(0)=0$~\endnote{As the warp factor is given by the difference $A_1\equiv A(y_1)-A(0)$, an arbitrary constant for $A(0)$ would not change the physical properties of the metric.}.
The Ricci scalar is
\begin{equation}
R = \frac{4 (5 - 2 \nu^2)}{\nu^4} \frac{1}{(y_s - y)^2} \,,  \label{eq:R}
\end{equation}
so that the naked singularity at $y = y_s$ becomes manifest.

\subsection{Black Hole/Brane Background}
\label{subsec:BH_background}

Let us now consider a generalization of the AdS-Schwarzschild solution, in particular the BH~\endnote{Black holes feature compact event horizons and a point-like singularity, whereas black branes are extended higher-dimensional objects whose event horizons stretch uniformly along one or more spatial directions, with the latter being higher-dimensional generalizations of the former. Nevertheless, in line with common practice in the field, in the present work we will refer to black brane solutions, as the one discussed in Section~\ref{subsec:BH_background}, as black hole solutions.} metric of the form~\endnote{Other kinds of BH solutions, called black strings, have been recently studied in Ref.~\cite{Fichet:2026ace} within the model of Equation~(\ref{eq:W_lineardilaton_nu}).}
\begin{equation}
ds_{\textrm{BH}}^2 = g_{MN} dx^M dx^N \equiv e^{-2 A(y)} \left(  h(y) dt^2 - d\vec{x}^2 \right) - \frac{dy^2}{h(y)} \,, \label{eq:dsBH}
\end{equation}
where $h(y)$ is a blackening factor which vanishes at the position of the event horizon, $y = y_h<y_s$. The EoM with this metric reads
\begin{eqnarray} 
&&\frac{h^{\prime\prime}}{h^\prime} = 4 A^\prime  \,, \label{eq:eomh1}\\
&&A^{\prime\prime}
= \frac{\kappa^2}{3} \phi^{\prime \, 2}  \,, \label{eq:eomh2}\\
&&A^{\prime\, 2} 
= \frac{h^\prime}{4 h} A^\prime - \frac{\kappa^2}{6} \frac{V(\phi)}{h} + \frac{\kappa^2}{12} \phi^{\prime\, 2} \,,  \label{eq:eomh3}\\
&&\phi^{\prime\prime}  = \left(  4 A^\prime - \frac{h^\prime}{h} \right) \phi^\prime + \frac{1}{h} V^\prime(\phi)  \,, \label{eq:eomh4}
\end{eqnarray}
where, to simplify the expressions, we have omitted the boundary terms. The solution turns out to be the same as in Equations~(\ref{eq:sol_A_T0}) and~(\ref{eq:sol_T0}) with the blackening factor
\begin{equation}
h(y) = 1  - \left( \frac{y_s - y_h}{y_s - y} \right)^{\frac{4 - \nu^2}{\nu^2}} \,. \label{eq:sol_T}
\end{equation}
Two of the integration constants have been fixed by the boundary conditions $\lim_{y \to -\infty}h(y) = 1$  and $h(y_h) = 0$. For $\nu<2$, the condition $0<h(y_1)<1$ requires $y_1<y_h<y_s$, which then excludes the singularity from the interval, as it is screened by the horizon. The Ricci scalar for the BH metric is
  \begin{equation}
R_{\textrm{BH}} = R  + \frac{3}{\nu^2}  \left( \frac{ y_s - y_h }{ y_s - y} \right)^{ \frac{4 - \nu^2}{\nu^2} } \frac{1}{(y_s - y)^2}  \,, 
  \end{equation}
  where $R$ is the Ricci scalar in the absence of BH, cf.~Equation~(\ref{eq:R}). It is clear from the above expressions that the case of no BH is recovered for $y_h\to y_s$.

\subsection{Thermodynamic Relations}
\label{sec:thermodynamics}
From Equation~(\ref{eq:dsBH}), the Hawking temperature $T_h$ and the entropy density $s_h$ of the BH can be expressed~as
\begin{eqnarray}
  T_h &=& \frac{1}{4\pi} e^{-A(y_h)} |h^\prime(y)|_{y = y_h} = \frac{1}{4\pi} \left( \frac{4}{\nu^2} - 1 \right) \frac{1}{y_s} \left( 1 - \frac{y_h}{y_s}  \right)^{\frac{1 - \nu^2}{\nu^2}}  \,, \\
  s_h &=& \frac{4\pi}{\kappa^2} e^{-3 A(y_h)}  = \frac{4\pi}{\kappa^2} \left( 1 - \frac{y_h}{y_s} \right)^{\frac{3}{\nu^2}}  \,.
\end{eqnarray}
By using these two expressions, we can write the entropy density as a function of $T_h$ as~follows
\begin{equation}
s_h(T_h) = \frac{4\pi}{\kappa^2} \left[ \frac{4\pi  \nu^2 y_s }{ 4 - \nu^2 }\right]^{\frac{3}{1 - \nu^2}}  T_h^{\frac{3}{1-\nu^2}} \,.
\end{equation}  

The quantity $T_h$ has a key role in the phase transition. To appreciate this, let us consider the case where the SM is either located in the IR brane or just propagating in the 5D bulk along with bulk gravitons and radions, as a thermal plasma with a temperature $T$. The temperature $T$ is not necessarily equal to the BH temperature, $T_h$. Then, the model would describe an out-of-equilibrium situation between both subsystems, for which the relevant thermodynamic relations for the internal and free energy densities are~\cite{Creminelli:2001th,Megias:2018sxv,Fichet:2026nct}
\begin{eqnarray}
\rho_h(T_h) &=& T_h \, s(T_h) - \int_0^{T_h} s(\hat T_h) \, d\hat T_h \,, \label{eq:rho} \\
  f_h(T_h,T) &=& (T_h - T) s(T_h) - \int_0^{T_h} \, s(\hat T_h) \, d\hat T_h \,. \label{eq:f}
\end{eqnarray}
In fact, by computing $\partial f_h(T_h,T) / \partial T_h$ from Equation~(\ref{eq:f}), it becomes manifest that $f_h(T_h,T)$ has a minimum at $T_h = T$ that amounts to
\begin{equation}
f_{\textrm{min}} = f_h(T) = - \int_0^T s(\hat T_h) d \hat T_h = - \frac{4\pi}{\kappa^2} \frac{1 - \nu^2}{4 - \nu^2} \left[ \frac{4\pi  \nu^2 y_s }{ 4 - \nu^2 }\right]^{\frac{3}{1 - \nu^2}}    T^\frac{ 4 - \nu^2}{ 1 - \nu^2}  \,,
\end{equation}
corresponding to the equilibrium situation. Notice that, upon time compactification with periodicity $1/T$, there appears a conical singularity that only cancels at the minimum at $T_h=T$, where the equilibrium is restored~\cite{Creminelli:2001th}.

The case $\nu=0$ reproduces the known results from Ref.~\cite{Creminelli:2001th}, for which the free energy decreases as $T^4$, as radiation. For the case $0<\nu<1$, it decreases faster than radiation, while for $1<\nu<2$, it increases with decreasing temperatures.
The case $\nu=1$, corresponding to the linear dilaton background, is ill-defined and will be worked out later on.

\section{The Linear Dilaton Background for Particle Physics}
\label{sec:particlephysics}

We will hereby consider the critical case $\nu=1$, where the superpotential and the bulk potential are 
\begin{equation}
W(\bar\phi) = \frac{6k}{\kappa^2}e^{\bar\phi} \,, \qquad V(\bar\phi) = -\frac{9k^2}{2\kappa^2}e^{2\bar\phi} \,,
\label{eq:W_lineardilaton}
\end{equation}
where the parameter $k\lesssim M_5$ determines the 5D curvature. Crucially, the superpotential in Equation~(\ref{eq:W_lineardilaton}) introduces a singularity at a finite proper coordinate value, \(y=y_s\), reproducing the characteristic behavior of soft wall setups. It leads to a gapped continuum spectrum, and the hierarchy problem is, more conventionally, solved in the same way as in RS theories, with fundamental scales $M_5$ and $k$, and a derived TeV scale after warping. The solution for the background in proper coordinates is (cf.~Equation~(\ref{eq:sol_T0})~\endnote{The solution of the EoM, Equation~(\ref{eq:phiA}), leads, in fact, to $A(y) = \bar\phi(y) + c$, where $c$ is a constant that can be fixed by choosing $A(0) = 0$ so that  $c = -\bar \phi(0) = -\bar v_0$. As for the case of an arbitrary $\nu$, the value of $c$ does not affect the warp factor $A_1$.})
\begin{equation}
\bar\phi(y)=-\log[k(y_s-y)] \,, \qquad A(y)= -\log\left( 1 - \frac{y}{y_s} \right) \,, \label{eq:phiy_Ay}
\end{equation}
and the BH blackening factor
\begin{equation}
h(y)=1-\left(\frac{y_s-y_h}{y_s-y} \right)^3  \,.
\end{equation}

We will also consider in this section two branes at $y=0$ (the UV boundary) and $y=y_1$ (the IR or Higgs brane), where we assumed the Standard Model and, in particular, the Higgs, to be located, such that the values of the IR brane location $y_1$ and the singularity $y_s$ will be dynamically determined by the brane potentials $\lambda_a(\phi)$, fixing the field $\bar\phi$ at the values $\bar v_0\equiv\kappa v_0/\sqrt{3} $ and $\bar v_1\equiv \kappa v_1/\sqrt{3}$ in the UV and IR branes, respectively, such that
\begin{equation}
ky_s=e^{-\bar v_0} \,, \qquad ky_1=e^{-\bar v_0}-e^{-\bar v_1} \,. \label{eq:kys_ky1}
\end{equation}
We can then write for the scalar field the background value $\bar\phi(y)=\bar v_0-\log(1-y/y_s)$.

Notice that the first expression in (\ref{eq:kys_ky1}) demands that $k y_s > 0$. Moreover, the solution of the hierarchy problem between $M_5$ and the TeV scale is achieved for a given value of the warp factor at the IR brane $A(y_1)\equiv A_1$, which imposes the relation
\begin{equation}
\bar v_1 - \bar v_0 = A_1 \,.
\end{equation}

As we will see, the squared mass gap of the continuum graviton and radion spectrum is then given by 
\begin{equation}
m_g^2 = \frac{9}{4} \eta^2 \,,
\end{equation}
with
\begin{equation}
\eta = \pm 1/y_s  \,.
\label{eq:rho2}
\end{equation}
Using the fact that $dz = \pm e^{A(y)} dy$, where $\pm$ corresponds to the sign of $\eta$, the relation between conformally flat and proper coordinates in the model of Equation~(\ref{eq:W_lineardilaton}) turns out to be 
\begin{equation}
\eta \cdot ( z - z_0)  = - \log( 1 - y/y_s) \,, \label{eq:z_y}
\end{equation}
and the background in these coordinates is given by
\begin{equation}
\bar\phi(z) =  A(z) + \bar v_0  \,, \qquad A(z) = \eta \cdot (z - z_0) \,, \label{eq:phiALD}
\end{equation} 
where we fix $z_0\equiv 1/k$. Therefore, in conformally flat coordinates, the dilaton $\phi$ is linear, and the corresponding model is dubbed the linear dilaton model (LDM).

According to Equation~(\ref{eq:rho2}), the parameter $\eta$ can have both signs, and accordingly, two classes of theories are implemented. The solution with a negative sign of the parameter $\eta$ in (\ref{eq:rho2}), i.e., $\eta=-1/y_s$, was considered in Refs.~\cite{Antoniadis:2011qw,Cox:2012ee}. Here, we will only consider the solution with $\eta=1/y_s$.

\subsection{The $\eta>0$ Case: The Continuum LDM}
\label{subsec:continuum_model}

In this section, we will consider the $\eta=1/y_s>0$ case in Equation~(\ref{eq:rho2})~\cite{Megias:2021mgj}. The IR brane location in conformally flat coordinates $z_1$ is dynamically determined, as well as the value of~$y_1$, by the IR fixing of the dilaton at the value $\bar v_1$. For simplicity, we will here fix $\bar v_1=0$ and $\bar v_0=-A_1<0$. Then, one finds from Equation~(\ref{eq:kys_ky1}) that $k y_s = e^{A_1}$ and $k(y_s - y_1) = 1$. 
The hierarchy problem is then solved by fixing $A_1$ such that $\eta=\mathcal O(\textrm{TeV})$, as given by Equation~(\ref{eq:rho2}), for $k\lesssim M_5$ with
\begin{equation}
\eta = k \, e^{-A_1} \,.
\end{equation}

The value of $M_5$ is determined by the relation between $M_5$ and $k$ with the 4D Planck scale. 
To evaluate the relationship between $M_5$ and $M_4$, we must consider that the kinetic term of the graviton zero mode has to be canonical. As we will see, the zero mode has a constant profile in proper coordinates, i.e., for the theory defined in the interval $[0,y_s]$
\begin{equation}
h^{(0)}_{\mu\nu}(x,y)=h_0\cdot h_{\mu\nu}(x) \quad  \Rightarrow\quad \int_0^{y_s}e^{-2A(y)}h_0^2 \, dy = 1\quad\Rightarrow\quad h_0=\sqrt{3\eta}  \,.  \label{eq:grav_norm}
\end{equation}
%
%where the factor of 2 in front of the integral comes from integration in the orbifold. 
We then obtain~\endnote{The expression in Equation~(42) is obtained by integrating the action in the extra dimension, i.e.,
  \begin{equation}
    \frac{1}{2\kappa^2} \times 3  \int_0^{y_s} e^{-2A(y)} dy = \frac{M_4^2}{2} \,, \tag{41}
  \end{equation}
  where we have identified the result in terms of the Planck scale in 4D; see, e.g., Equation~(\ref{eq:grav-lagrangian}) below. The factor of $3$ in front of the integral comes from the normalization of the graviton zero mode, i.e., $h_0^2 = 3 \eta$ where $\eta = 1/y_s$.}
\begin{equation}\setcounter{equation}{42}
\kappa^2 M_{\rm 4}^2=3\int_0^{y_s}e^{-2A(y)}dy \quad \Rightarrow \quad M_5^3=\eta M_{\rm 4}^2 \,, \label{eq:Mpl}
 \end{equation}
which yields, for $\eta=\mathcal O(\textrm{TeV})$, $M_5\simeq 10^{13}\textrm{ GeV}\simeq10^{-5}M_{4}$, and correspondingly, a warp factor $A_1\simeq 23$. In conformal coordinates, the location of the branes in units of $\eta$ are at $\eta z_0 = e^{-A_1}$ and $\eta z_1 = A_1 + e^{-A_1}\simeq A_1$, while the singularity is located at infinity, $z_s\to\infty$. The length of the fundamental region is $y_1\simeq y_s\simeq 10^{-17}$ cm, much larger than the corresponding one in the RS model $\sim 10^{-31}$ cm, but still much smaller than that in ADD theories~\cite{Arkani-Hamed:1998jmv}~$\sim 10^{-2}$ cm (for the case of two extra dimensions), where only gravity propagates in the bulk.

We can, in principle, consider two kinds of fundamental intervals for our theory:

\begin{itemize}
\item
When the fundamental interval is $[0,y_1]$, i.e.,~$[z_0,z_1]$, in conformal coordinates, the theory spectrum is discrete, with the first mode mass being $\mathcal O(\eta)$.
\item
When the fundamental interval is $[0,y_s]$, i.e.,~$[z_0,\infty)$, in conformal coordinates (as we will consider here), this theory predicts a continuum spectrum with an $\mathcal O(\eta)$ gap: the \textit{continuum linear dilaton model} (CLDM). This is the interval we will consider in this review, where we focus on the case $\eta>0$. Notice that this makes the main difference with respect to the $\eta<0$ case considered in Ref.~\cite{Antoniadis:2011qw}, in which gravity decouples ($M_4\to\infty$) when considering the infinite interval, $z_s\to\infty$.
\end{itemize}

It is worth emphasizing that, in this theory, the warp factor solves the hierarchy problem between the intermediate scale $M_5$ and the TeV scale. This means that the UV completion of this theory should be a string theory, with the string mass at the intermediate scale, $M_s\simeq M_5$, thus solving the hierarchy problem between the Planck scale and $M_5$. 

\subsection{The Gauge Bosons}
\label{subsec:gauge_bosons}

As we will see in this section, this class of background does not support gauge bosons propagating in the bulk of the extra dimension, unless the gauge bosons' KK modes are in the 100 TeV range.
%so that we will consider the whole SM localized at the IR brane. %Notice that in this setup, as in RS models, $M_5$ is a fundamental scale, while the TeV scale and the 4D Planck mass are derived from the theory warp factor.

To begin with, let us assume that only the Higgs is localized on the IR brane, while gauge bosons are propagating in the bulk. We will also assume a custodial model based on the bulk gauge group~\cite{Agashe:2003zs,Carena:2018cow}
\be
SU(3)_c\otimes SU(2)_L\otimes SU(2)_R\otimes U(1)_X \,,
\ee
where $X\equiv B-L$,
with 5D gauge bosons $(\mathcal G,W_L,W_R,X)$, and 5D couplings $(g_c,g_L,g_R,g_X)$~\endnote{The 5D ($g_5$) and 4D ($g_4$) couplings are related by $g_4 = g_5/\sqrt{y_s}$.}, respectively. 

The breaking $SU(2)_R\otimes U(1)_X \to U(1)_Y$, where $Y$ is the SM hypercharge with gauge boson $B$ and coupling $g_Y$, is carried out in the UV brane by boundary conditions. Therefore, the gauge fields $(W_L^a,W_R^a,X)$ define $(W_L^{a},W_R^{1,2},B,Z_R)$, with (UV and IR) boundary conditions, as
\begin{align}
&W_L^a\ (a=1,2,3)\,, & (+,+)   \label{eq:Wpp}\\
B&=\frac{g_X W_R^3+g_RX}{\sqrt{g_R^2+g_X^2}}\,,&(+,+) \label{eq:B}\\
&W_R^{1,2}\,, & (-,+) \label{eq:WR}\\
Z_R&=\frac{g_R W_R^3-g_X X}{\sqrt{g_R^2+g_X^2}}\,. &(-,+)  \label{eq:Zmp}
\end{align}
The $SU(2)_L\otimes SU(2)_R$ symmetry is unbroken in the IR brane, where all composite states are localized, such that the \textit{custodial} symmetry is exact. After electroweak breaking, the different fields are denoted by $(W_L^\pm,W_R^\pm,Z_L,Z_R,A_L)$.

The covariant derivative for fermions is
\be
\slashed{D}=\slashed\partial-i\left[g_L\sum_{a=1}^3 \slashed{W}_L^{a} T_L^a+g_R\sum_{b=1}^2 \slashed{W}_R^{b} T_R^b+
g_Y \slashed{B}\,Y+g_{Z_R}\slashed{Z}_R\,Q_{Z_R}
\right] \,,
\ee
where $g_Y$ and $g_{Z_R}$ are defined in terms of $g_R$ and $g_X$ as
\be
g_Y=\frac{g_Rg_X}{\sqrt{g_R^2+g_X^2}}\,, \qquad g_{Z_R}=\sqrt{g_R^2+g_X^2}\ ,
\ee
and the hypercharge $Y$ and the charge $Q_{Z_R}$ are defined by
\be
Y=T_R^3+Q_X\,, \qquad Q_{Z_R}=\frac{g_R^2T_R^3-g_X^2 Q_X}{g_R^2+g_X^2} \,,
\ee
with $Q_X=(B-L)/2$.

Electroweak symmetry breaking is triggered in the IR brane by the bulk Higgs \mbox{bi-doublet}
\be
\mathcal H=\left( \begin{array}{cc}
H^{0}_2 & H_1^+ \\
H_2^- & H_1^0
\end{array}\right)\,, \qquad Q_X=0  \,,
\ee
where the rows transform under $SU(2)_L$ and the columns under $SU(2)_R$. We will denote their VEVs as $\langle H_2^0\rangle\equiv v_2/\sqrt{2}$ and $\langle H_1^0\rangle \equiv v_1/\sqrt{2}$ so that we will introduce the angle $\beta$ as $\cos\beta=v_1/v$ and $\sin\beta=v_2/v$, with $v=\sqrt{v_1^2+v_2^2}$. %We will find it useful to add an extra Higgs bi-doublet 

\subsubsection{Massless Gauge Bosons}
\label{subsec:massless_gaugebosons}

The Lagrangian is~\endnote{We are using, in this section, the gauge $A_5=0$.}
\be
\mathcal L= \int_0^{y_s} dy\left[ -\frac{1}{4}\tr F_{\mu\nu}F^{\mu\nu}-\frac{1}{2}e^{-2A}\tr A'_\mu A'_\mu  \right] \,,
\label{eq:Lagrangian_GB}
\ee
where the trace is over the gauge indices. After Fourier transforming the coordinates $x^\mu$ into momenta $p^\mu$, we can make the field decomposition
$A_\mu(p,y)=f_A(y) A_\mu(p)/\sqrt{y_s}$, and
the EoM of the fluctuations is given by~\cite{Cabrer:2010si}
\begin{equation}
p^2 f_A + \frac{d}{dy}(e^{-2A} f_A^\prime(y)) = 0 \,.  \label{eq:fAy}
\end{equation}
In conformal coordinates, and after rescaling the field by ${ f}_A(z) = e^{A(z)/2} \hat f_A(z)$, we obtain the Schr\"odinger like form for the equation of motion
\begin{equation}
-\ddot{{\hat f}}_A(z) + V_A(z) {\hat f_A}(z) = p^2 {\hat f_A}(z) \,, \label{eq:ftA}
\end{equation}
where $\dt{X}\equiv{dX}/dz$, and the effective Schr\"odinger potential is
\begin{equation}
V_A(z) =  \frac{1}{4} {\dt A}^2(z) - \frac{1}{2} {\ddot A}(z)  \,. \label{eq:VA}
\end{equation}
Plugging Equation~(\ref{eq:phiALD}) into this equation, we find the following result for the effective potential
%
%\begin{equation}
%V_A(z) =  \left\{ 
%\begin{array}{cc}
%\frac{3}{4z^2}  & \qquad z  \le   z_1  \\
%\frac{\rho^2}{4} &  \qquad   z_1 < z  
%\end{array} \,, \right. \label{eq:VAz}
%\end{equation}
%
\begin{equation}
V_A(z) = \frac{\eta^2}{4} \,. \label{eq:VAz}
\end{equation}
%where $\eta$ is defined in Equation~(\ref{eq:rho2}). %~\endnote{Note that $V_A(z)$ is discontinuous at $z=z_1$, and given Equation~(\ref{eq:ftA}) this induces a discontinuity in $\ddt{{\hat f}}_A(z_1)$. Nevertheless $\hat f_A(z)$ and $\dt{{\hat f}}_A(z)$ are continuous functions.}. We can see that in the IR regime the potential is constant
%%
%\begin{equation}
%V_A(z) \stackrel[z > z_1]{=}{} \frac{1}{4} \rho^2  \,.
%\end{equation}
%%
Then, we find the existence of a mass gap in the potential, which translates into a gap followed by a continuum KK spectrum of gauge bosons.

\begin{itemize}
\item {\it Green's functions:} We will compute the Green's functions for gauge bosons propagating in the bulk of the 5D spacetime from $y$ to $y^\prime$, where both $y$ and $y^\prime$ are considered arbitrary. To compute the Green's function, we have to solve an inhomogeneous version of the equation of motion, Equation~(\ref{eq:fAy}). This is given by
\begin{equation}
p^2 G_A(y,y^\prime;p) + \frac{d}{dy}\left( e^{-2A} G_A^\prime(y,y^\prime;p) \right) = \delta(y-y^\prime) \,,  \label{eq:GAy}
\end{equation}
where the prime in $G_A^\prime$ denotes a derivative with respect to the variable~$y$. One should start from the general solution of this equation and choose the integration constants such that the corresponding boundary conditions are fulfilled. The Green's functions are subject to the following boundary and matching conditions
\begin{eqnarray}
\Delta G_A(y^\prime) = 0 \,, \qquad \Delta G_A^\prime(y^\prime) = e^{2 A(y^\prime)} \,, \qquad \Delta G_A(y_1) = 0 \,, \qquad  \Delta G_A^\prime(y_1) = 0 \,, \label{eq:GA_bc}
\end{eqnarray}
in addition to boundary conditions in the UV brane, which will be discussed next. In these expressions, only the behavior on the first variable $y$ is shown in the Green's functions, and we define the jump function as $\Delta f(y) \equiv \lim_{\epsilon \to 0}\left( f(y + \epsilon)  - f(y - \epsilon)\right)$.

\item {\it Gauge bosons with Neumann and Dirichlet boundary conditions:} While the SM photon ($A_L$) is subject to the Neumann boundary condition in the UV brane, in the considered extension of the SM, we will consider Dirichlet boundary conditions on this brane for the extra massless gauge bosons ($W_R^\pm,Z_R$). Then, the boundary conditions in the UV brane turn out to be
  \begin{equation}
   G_A^{(++) \; \prime}(y_0) = 0  \,, \qquad G_A^{(-+)}(y_0) = 0 \,, \label{eq:ND_bc}
  \end{equation}
  for the Green's functions $G_A^{(++)}$ and $G_A^{(-+)}$, respectively. These conditions are supplemented by the other conditions in Equation~(\ref{eq:GA_bc}). %As an example, the integration constants in the domain $y \le y^\prime$ turn out to fulfill the relations
%\begin{equation}
%C_1^I = - \frac{Y_0\left( \frac{p}{k} \right)}{J_0\left( \frac{p}{k} \right)}  C_2^I  \quad \textrm{(Neumann)} \,, \qquad C_1^I = - \frac{Y_1\left( \frac{p}{k} \right)}{J_1\left( \frac{p}{k} \right)}  C_2^I \quad \textrm{(Dirichlet)} \,.
%\end{equation}
%
The results for the IR-brane-to-IR-brane Green's functions for massless gauge bosons for both $(+,+)$ and $(-,+)$ boundary conditions turn out to be
  \begin{eqnarray}
    G_A^{(+,+)}(y_1,y_1;p) &=&  \left[  \Delta_A^- \left( \frac{\eta}{k}\right)^{\delta_A}  - \Delta_A^+  \right] \frac{\Delta_A^-}{\delta_A}   \frac{k}{4 p^2}  \,, \qquad (A_L) \,, \label{eq:GApp} \\
    G_A^{(-,+)}(y_1,y_1;p) &=& \left[  \left( \frac{\eta}{k} \right)^{\delta_A} - 1 \right] \frac{1}{\delta_A} \frac{k}{\eta^2} \,,  \qquad\qquad (W_R^\pm, Z_R) \,, \label{eq:GAmp}
  \end{eqnarray}
  where
  \begin{equation}
  \Delta_A^\pm \equiv \pm \delta_A - 1 \,, \qquad \delta_A = \sqrt{1 - \frac{4 p^2}{\eta^2}} \,.
  \end{equation}
  We have indicated in Equations~(\ref{eq:GApp}) and (\ref{eq:GAmp}) the field(s) to which each propagator~\mbox{applies}.

\end{itemize}

\subsubsection{Standard Model Massive Gauge Bosons}
As we are considering the Higgs sector localized on the IR brane, in the case of the SM massive gauge bosons $A_\mu$ (i.e.,~$W_L$ and  $Z_L$),  there are extra terms in the 5D Lagrangian, Equation~(\ref{eq:Lagrangian_GB}), as%as~\cite{Cabrer:2011fb}
\be
\Delta\mathcal L_5=\left(-\frac{1}{2}  M_Z^2 Z_\mu^2-  M_W^2 |W_\mu|^2 \right) \delta(y-y_1)\,, \qquad M_A^2 = y_s m_A^2 \,,  \label{eq:Delta_L5}
\ee
(for $A=Z_L,W_L$), which leads to a modification of the equation of motion for the Green's function, as
\be
\left[ p^2- y_s m_A^2\,\delta(y-y_1) \right] G_{A,M}(y,y^\prime;p)  + \frac{d}{dy}\left( e^{-2A} G_{A,M}^\prime(y,y^\prime;p) \right) = \delta(y-y^\prime) \,.  \label{eq:GAymassive}
\ee
Using Equation~(\ref{eq:GAymassive}), the derivative of the Green's functions turns out to be discontinuous at $y = y_{1}$, with a jump given by~\endnote{In the following, we will assume that $y^\prime \ne y_{1}$. Then, $\int_{y_{1} - \epsilon}^{y_{1} + \epsilon} dy \, \delta(y-y^\prime) = 0$, and the right-hand side of Equation~(\ref{eq:GAymassive}) does not contribute to Equation~(\ref{eq:GAmassivejump}). In this way, we compute the Green's functions $G_{A,M}(y,y_1)=\lim_{y^\prime\to y_1}G_{A,M}(y,y^\prime)$. Alternatively, we could directly compute the Green's function $G_{A,M}(y,y_1)$ with a jump at $y=y_1$ given by $$\Delta G_{A,M}^\prime(y_1,y_1) = m_A^2 y_s \, e^{2A(y_{1})} G_{A,M}(y_{1},y_1)+e^{2A(y_{1})}.$$ It is possible to check that both procedures lead to the same result.}
\begin{equation}
\Delta G_{A,M}^\prime(y_1,y^\prime) = m_A^2 y_s \, e^{2A(y_{1})} G_{A,M}(y_{1},y^\prime) \,.  \label{eq:GAmassivejump}
\end{equation}

Finally, the results for the IR-brane-to-IR-brane Green's functions for massive gauge bosons for both $(+,+)$ and $(-,+)$ boundary conditions turn out to be
  \begin{eqnarray}
    G_A^{(+,+)}(y_1,y_1;m_A;p) &=&   \frac{ \Delta_A^- \left( \frac{\eta}{k} \right)^{\delta_A}  - \Delta_A^+  }{ \delta_A - k y_s \left( \frac{m_A}{\eta} \right)^2 \left[ \frac{\Delta_A^-}{\Delta_A^+}  \left( \frac{\eta}{k}  \right)^{\delta_A} - 1 \right]  }  \Delta_A^- \frac{k}{4 p^2}  \,,  \quad (W_L^\pm, Z_L) \,, \label{eq:GApp_mA} \\
    G_A^{(-,+)}(y_1,y_1;m_A;p) &=&  \frac{ \left( \frac{\eta}{k} \right)^{\delta_A} - 1}{ \delta_A - k y_s \left( \frac{m_A}{\eta} \right)^2 \left[ \left( \frac{\eta}{k} \right)^{\delta_A} - 1\right] } \frac{k}{\eta^2} \,. \label{eq:GAmp_mA}
  \end{eqnarray}
  Notice that these results reproduce, in the limit $(m_A \to 0)$, the Green's functions for massless gauge bosons provided in Section~\ref{subsec:massless_gaugebosons}.  Notice that the SM massive gauge bosons $(W_L^\pm, Z_L)$  have $(+,+)$ boundary conditions. For completeness, we also enclose the $(-,+)$ boundary conditions for other hypothetical massive gauge fields in the custodial model, although none of these fields are present in our model.

We summarize in Table~\ref{tab:BC} the boundary conditions of the gauge fields of the custodial~\mbox{model}.

\begin{table}[H]
 % \centering
 \caption{Summary of the boundary conditions for the gauge fields in the custodial model.}
\label{tab:BC}
\begin{tabularx}{\textwidth}{LCC}
    \toprule
     & $(+,+)$ 
     & $(-,+)$ \\
    \midrule
     { Massless }  &   $A_L$  & $W_R^\pm$, $Z_R$ \\
   \midrule
        { Massive}  &  $W_L^\pm$, $Z_L$ &  \rule[0.7mm]{0.5cm}{0.8pt}  \\
    \bottomrule
  \end{tabularx}

\end{table}

\subsubsection{Electroweak Precision Observables}
When the electroweak symmetry is broken, there is a mixing between the SM fields $W_L$ and $Z_L$ and the heavy modes of $W_{L,R}$ and $Z_{L,R}$ induced by the Lagrangian
\be
\mathcal L=\tr|g_L^5 W_L^aT_L^a \mathcal H-g_R^5 \mathcal H W_R^a T_R^a |^2 \,,
\label{eq:mixing}
\ee
where we indicate with the script $g_5$ the 5D gauge couplings, related to the 4D couplings $g_4$ by $g_5 = g_4 \sqrt{y_s}$.

After putting the Higgs bidoublet $\mathcal H$, which we assume to be localized on the IR brane,
\be
\mathcal H=\begin{pmatrix} H_2^0 & H_1^-\\
H_2^- & H_1^0
\end{pmatrix}
\ee
at its minimum, $\langle H_{1,2}^0\rangle=v_{1,2}$, with $v_1^2+v_2^2=v^2$ and $v=246.22$ GeV, and the Lagrangian (\ref{eq:mixing}) gives rise to the quadratic terms
%
%\begin{align}
%\mathcal L=&\frac{v^2}{4} y_s \bigg[ g_L^2 W_L(z_1,x) W_L(z_1,x)-g_L g_R W_L(z_1,x) W_R(z_1,x)\nonumber\\
%+&\frac{g_L^2}{c_L^2}Z_L(z_1,x)Z_L(z_1,x)-g_L g_R \frac{c_R}{c_L}Z_L(z_1,x)Z_R(z_1,x) \bigg] \,.
%\end{align}
%
%
\begingroup
\makeatletter\def\f@size{8.7}\check@mathfonts
\def\mketag@@@#1{\hbox{\m@th\normalsize\normalfont#1}}%
\begin{align}
&\mathcal L = \frac{v^2}{4} y_s \bigg[ g_L^2 W_L(z_1,x) W_L(z_1,x) + g_R^2 W_R(z_1,x) W_R(z_1,x) - \frac{2 v_1 v_2}{v^2} g_L g_R W_L(z_1,x) W_R(z_1,x)  \nonumber\\
&+\frac{1}{2}\frac{g_L^2}{c_L^2}Z_L(z_1,x)Z_L(z_1,x) +  \frac{1}{2}g_R^2c_R^2Z_R(z_1,x)Z_R(z_1,x)  - g_L g_R \frac{c_R}{c_L}Z_L(z_1,x)Z_R(z_1,x) \bigg] \,,
\end{align}
\endgroup
where $W_X W_X \equiv W_X^- W_X^+$ for $X = L, R$, and $W_L W_R \equiv W_L^- W_R^+ + W_L^+ W_R^-$. One has $v_1 = v \cdot \cos\beta$ and $v_2 = v \cdot \sin\beta$, and then $2 v_1 v_2 / v^2 = 2 t_\beta / (1 + t_\beta^2)$, where we have defined $t_\beta \equiv \tan\beta$. In the custodial limit,  $t_\beta = 1$ and $v_1 = v_2 = v/\sqrt{2}$.

The fields $W_L$ and $Z_L$ have a zero mode, which is the corresponding SM field, and a gapped continuum of states, while the fields $W_R$ and $Z_R$ do not possess a zero mode and only have the continuum of states after the mass gap. For the electroweak observables contributing to the new physics, the oblique $T$, $S$ and $U$ parameters are defined as~\cite{Cabrer:2011fb}
\begin{equation}
  \begin{split}
\alpha T&=\frac{\Pi_{WW}(0)}{m_W^2}-\frac{\Pi_{ZZ}(0)}{m_Z^2} \,,\\
\alpha S&= 4 s_L^2 c_L^2 \Pi^\prime_{ZZ}(0) \,,  \\
\alpha (S + U) &= 4 s_L^2 \Pi^\prime_{WW}(0) \,. 
 \end{split}
\end{equation}

For the computation of these parameters, we need to select, as external fields, the zero modes of either $W_L$ or $Z_L$, and only propagate the continuum of states. We will then define the Green's functions propagating only the continuum of states as
\begin{equation}
 \begin{split}
\mathcal G_{W_L,Z_L}(p;z_1,z_1)&=G_{W_L,Z_L}(p;z_1,z_1)-G^0_{W_L,Z_L}(p) \,, \\
\mathcal G_{W_R,Z_R}(p;z_1,z_1)&=G_{W_R,Z_R}(p;z_1,z_1) \,.
 \end{split}
 \label{eq:GLRp}
\end{equation}

\begin{figure}[H]
%\begin{center}
 \includegraphics[width=0.5\textwidth]{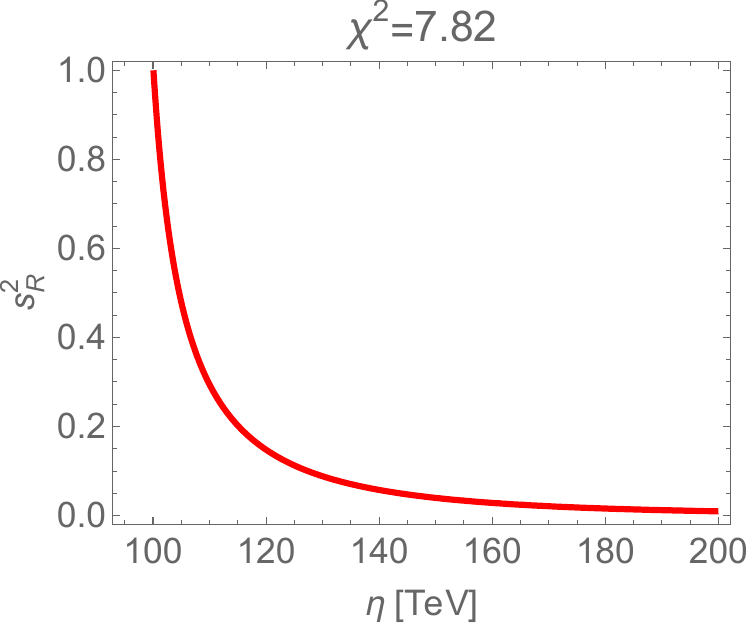}   
%\end{center}
     \caption{$\chi^2$ distribution for the $S$, $T$ and $U$ observables. The line $\chi^2=7.82$ provides the 95\% CL lower bound on the parameter $\eta$.}
     \label{fig:chi2}
 \end{figure}

By the notation $\mathcal G(0)\equiv \mathcal G(0;z_1,z_1)$, a straightforward calculation yields
\be
\alpha T=m_W^2 y_s\left[\mathcal G_{W_L}(0)+ \frac{4 t_\beta^2}{(1+t_\beta^2)^2} \frac{g_R^2}{g_L^2}\mathcal G_{W_R} (0) \right] - m_Z^2 y_s\left[\mathcal G_{Z_L}(0)+\frac{g_R^2}{g_L^2}c_L^2 c_R^2\mathcal G_{Z_R} (0) \right] \,,
\ee
where $t_\beta=v_2/v_1$. Using the results of previous sections, we find
\begin{equation}
  \begin{split}
\mathcal G_{W_L} (0)&=\mathcal G_{Z_L}(0)=  - y_s ( k y_s - 2 \log(k y_s)) \equiv \mathcal G_L(0) \,, \\
\mathcal G_{W_R}(0) &=\mathcal G_{Z_R}(0)= - y_s (k y_s - 1)  \equiv \mathcal G_R(0) \,,
  \end{split}
  \label{eq:GLR0}
\end{equation}
so that
\be
\alpha T=m_W^2 y_s\frac{s_L^2}{c_L^2}\left[ \left( 1-\frac{1}{s_R^2}\frac{(1-t_\beta^2)^2}{(1+t_\beta^2)^2} \right) \mathcal G_R(0)-\mathcal G_L(0)\right] \,.
\ee
A similar calculation yields
\begin{align}
\alpha S &= 4 m_Z^4 y_s s_L^2 c_L^2  \left[  \mathcal G^\prime_L(0)+\frac{s_L^2 c_R^2}{s_R^2} \mathcal G^\prime_R(0) \right] \,, \\
\alpha U &= 4 m_Z^4 y_s  s_L^4 c_L^2 \left[ \left( 1-\frac{1}{s_R^2}\frac{(1-t_\beta^2)^2}{(1+t_\beta^2)^2} \right) \mathcal G^\prime_R(0) - \mathcal G^\prime_L(0) \right] \,,
\end{align}
where the prime stands for $\frac{d}{dp^2}$, and $\mathcal G'_{W_L}(0)=\mathcal G^\prime_{Z_L}(0)\equiv \mathcal G^\prime_L(0)$, $\mathcal G'_{W_R}(0)=\mathcal G^\prime_{Z_R}(0)\equiv \mathcal G^\prime_R(0)$. Then, we find
\begin{equation}
\begin{split}
\mathcal G^\prime_L(0)&= -2y_s^3 \left( k y_s - \log^2(k y_s) - \log(k y_s) - 1/2 \right) \,, \\
\mathcal G^\prime_R(0) &= -2 y_s^3 \left(  k y_s - \log(k y_s) - 1 \right)  \,.
\end{split}
\label{eq:GLRp0}
\end{equation}
%
%so that
%
%\begin{align}
%\alpha S &= s_L^2 c_L^2 \frac{m_Z^4}{\rho^4} \left[\frac{-40 (k y_s)^3+324 (k y_s)^2-977 (k y_s)+648 }{32 (k y_s)^2}-\frac{s_L^2 c_R^2}{s_R^2}  \frac{5 (k y_s)}{4}\right] \,, \\
%\alpha (S + U) &= s_L^2 c_L^4 \frac{m_Z^4}{\rho^4} \left[\frac{-40 (k y_s)^3+324 (k y_s)^2-977 (k y_s)+648 }{32 (k y_s)^2} - \frac{s_L^2}{c_L^2 c_R^2} \left(\frac{g_R^2}{g_Y^2}-1 \right) \frac{5 (k y_s)}{4}\right] \,, \\
%\alpha U &= -s_L^4 c_L^2 \frac{m_Z^4}{\rho^4} \left[ \frac{ 324 (k y_s)^2-977 (k y_s)+648}{32 (k y_s)^2} \right] \,. 
%\end{align}
%
We can see that, in the limit of a large value of $k y_s=e^{A_1}$, $\mathcal G_{L,R}(0), \mathcal G_{L,R}^\prime(0)=\mathcal O(k y_s)$, while 
$\mathcal G_R(0)-\mathcal G_L(0) \,, \; \mathcal G_R^\prime(0)-\mathcal G_L^\prime(0)=\mathcal O(k y_s)^0$, which is the cancelation that appears on the observables $T$ and $U$ in the custodial limit $t_\beta=1$. In the custodial limit, the observables $T$ and $U$ can be written as
\begin{equation}
\begin{split}
\alpha T&\simeq -2\left(\frac{m_W}{\eta}\right)^2 t_L^2 A_1 \,,\\
\alpha U&\simeq -8\left(\frac{m_Z}{\eta}\right)^4 s_L^4 c_L^2 A_1^2 \,,
\end{split}
\label{eq:TU}
\end{equation}
where the custodial cancellation is explicit.
However, the observable $S$ still gives a sizable contribution~as
\be
\alpha S\simeq -8\left( \frac{m_Z}{\eta} \right)^4s_L^2 c_L^2\left(1+\frac{s_L^2 c_R^2}{s_R^2} \right)e^{A_1} \,, \label{eq:alphaS}
\ee
which should be compared with the experimental values given in Ref.~\cite{ParticleDataGroup:2024cfk}
\begin{align}
S_{\rm exp} &= 0.018\,,\qquad \hspace{0.50cm} \sigma_S = 0.095 \,, \nonumber \\
T_{\rm exp} &= 0.04 \,, \qquad \hspace{0.65cm} \sigma_T = 0.12 \,, \\
U_{\rm exp} &= -0.013 \,,\qquad \hspace{0.15cm} \sigma_U = 0.086 \,, \nonumber
\end{align}
where we use the notation $X=X_{\rm exp}\pm \sigma_X$ and correlation coefficients $\rho_{ii}=1$, $\rho_{ij}=\rho_{ji}$ with
\be
\rho_{ST} = 0.91 \,,\qquad \rho_{SU} = -0.66 \,,\qquad \rho_{TU} = -0.88 \,.
\ee
Using the $3\times 3$ covariance $C$ matrix, $(C)_{ij}=\rho_{ij}\sigma_i\sigma_j$, one can build the $\chi^2$ distribution~as
\be
\chi^2= V\cdot C^{-1}\cdot V^T \,,\qquad V=(S_{\rm exp}-S,T_{\rm exp}-T,U_{\rm exp}-U)\,.
\ee
For $k\simeq M_5$, we can write the warp factor as $e^{A_1}\simeq (M_4/\eta)^{2/3}$. Considering the value of $\alpha(\eta)$ as that given by the one-loop running
\be
\frac{1}{\alpha(\eta)}=\frac{1}{\alpha(m_Z)}-\frac{16}{3\pi}\log\frac{\eta}{m_Z} \,,
\ee
We plot, in Figure~\ref{fig:chi2}, the $\chi^2$ distribution in the plane $(\eta,s_R^2)$. We can see that, independently of the value of the angle $s_R$, the absolute 95\% CL lower bound on $\eta$ is $\sim$100 TeV. This result then excludes, for TeV values of $\eta$, the SM propagating in the bulk and motivates the localization of the SM on the IR brane. Thus, states that propagate in the bulk are the graviton and the radion, apart from possible SM singlets.

\subsection{The Graviton Propagator}
\label{sec:graviton}

The graviton is a transverse traceless fluctuation of the metric of the form
\begin{equation}
ds^2 = e^{-2A(y)} g_{\mu\nu} \,dx^\mu dx^\nu - dy^2 \,, \qquad 
g_{\mu\nu}=\eta_{\mu\nu} + 2 \kappa\, h_{\mu\nu}(x,y) \,, \label{eq:ds2_graviton}
\end{equation}
where $h_\mu^\mu = \partial_\mu h^{\mu\nu} = 0$. The on-shell quadratic action, in the interval, is given by
\begin{equation}
\mathcal S =-\frac{1}{2}\int d^4x\int_0^{y_s} dy \, e^{-2A}\left[\partial_\rho h_{\mu\nu}\partial^\rho h^{\mu\nu}+e^{-2A}h'_{\mu\nu}h^{\prime \mu\nu}
\right] \,.
\label{eq:grav-lagrangian}
\end{equation}
We will use the ansatz $h_{\mu\nu}(x,y) = h(y) h_{\mu\nu}(x)$ from where the EoM can be written as
\begin{equation}
e^{2A(y)} \left( e^{-4A(y)} h^\prime(y) \right)^\prime+p^2 h(y) = 0 \,. \label{eq:eomgraviton1}
\end{equation}

In conformal coordinates, cf.~Equation~(\ref{eq:z_y}), and after rescaling the field by $h(z) = e^{3 A(z)/2} \tilde h(z)$, the EoM for the fluctuation can be written in the Schr\"odinger like form~\cite{Cabrer:2009we} as
\begin{equation}
-\tilde h^{\prime\prime}(z) + V_{h}(z) {\tilde h}(z) = p^2 {\tilde h}(z) \,,
\end{equation}
where the potential is given by
\begin{equation}
V_h(z) = \frac{9}{4} {A^\prime}^2(z) - \frac{3}{2} {A^{\prime\prime}}(z)  \,. \label{eq:Vh}
\end{equation}
An explicit evaluation of this potential with the expression of $A(z)$ given by Equation~(\ref{eq:phiALD}) leads to a constant value $V_h(z) = m_g^2$, where $m_g = 3\eta/2$ is the mass gap for the graviton, typical of a continuum of states with a gap.

The interaction with matter localized at the  brane $y=y_a$ is found as
\begin{equation}
\delta\mathcal  S=\frac{1}{2}\int d^5x\, \sqrt{-g}  \,  T^{\mu\nu}\,\delta g_{\mu\nu}\quad\Rightarrow\quad    \mathcal L_{5D}=\frac{1}{M_5^{3/2}} T^{\mu\nu}(x) h_{\mu\nu}(x,y_a) \delta(y - y_a) \,,  \label{eq:int_matter}
\end{equation}
 where the energy-momentum tensor is defined as
\begin{equation}
T^{\mu\nu}=\left.\frac{2}{\sqrt{-g}}\frac{\delta(\sqrt{-g}\mathcal L_{5D})}{\delta g_{\mu\nu}}\right|_{\eta_{\mu\nu}} \,,
\end{equation}
and so it is symmetric by definition.

%\subsection{Green's functions for the graviton}
%\label{subsec:GF_graviton}

The Green's function for $h_{\mu\nu}(x,y)$ in the transverse, traceless gauge, is given by~\cite{Hagiwara:2008jb}
%%
%\begin{equation}
%D_{\mu\nu,\rho\sigma}=\frac{1}{2}\left(\eta_{\mu\rho}\eta_{\nu\sigma}+\eta_{\mu\sigma}\eta_{\nu\rho}-\frac{2}{3}\eta_{\mu\nu}\eta_{\rho\sigma} \right) 
%G_h(y,y';p) \,,
%\end{equation}
%
\begin{equation}
D_{\mu\nu,\rho\sigma} =  G_h(y,y';p) P_{\mu\nu,\rho\sigma}(p) \,,
\end{equation}
where
\begin{equation}
P_{\mu\nu,\rho\sigma}(p) = \frac{1}{2}\left(\Theta_{\mu\rho}\Theta_{\nu\sigma} + \Theta_{\mu\sigma} \Theta_{\nu\rho} \right) -\frac{1}{3}\Theta_{\mu\nu}\Theta_{\rho\sigma}  \,, \qquad \Theta_{\mu\nu} = \eta_{\mu\nu} - \frac{p_\mu p_\nu}{p^2} \,.
\label{eq:P}
\end{equation}
We are using, for the 5D Green's function $G_h(y,y';p)$, a mixed representation where the 4D coordinates $x^\mu$ have been Fourier-transformed into 4D momenta $p^\mu$. After fixing the value of $y^\prime$, one can see that the Green's function obeys the same EoM as the field $h(y)$, except for an inhomogeneous Dirac delta term, i.e.,
\begin{equation}
p^2 G_h(y,y^\prime) + e^{2A(y)} \frac{d}{dy}\left( e^{-4A(y)} G_h^\prime(y,y^\prime) \right) =   e^{2A(y^\prime)} \delta(y-y^\prime) \,,  \label{eq:Ghy}
\end{equation}
where the prime indicates a derivative with respect to the variable~$y$, and, for simplicity, we omit the $p$ dependence from the argument of the Green's function. After substituting the explicit expression for the background $A(y)$, we find that the equation reads as
\begin{equation}
G_h^{\prime\prime}(y,y')  - \frac{4}{y_s - y} G_h^{\prime}(y,y') + \frac{1}{(y_s - y)^2}\left( \frac{p}{\eta} \right)^2  G_h(y,y') =  e^{4 A(y^\prime)}  \delta(y-y^\prime)  \,,  \label{eq:eom_graviton3}
\end{equation}
whose general solution is given in Ref.~\cite{Megias:2021mgj}. For the particular case where $y\leq y'$, we find the solution
\begin{eqnarray}
\hspace{-0.5cm} G_h(y,y^\prime) = \frac{1}{3\eta} \frac{1}{\delta_h} (1 - \bar y')^{\frac{3}{2}\Delta_h^+} \left(  - (1 - \bar y)^{\frac{3}{2}\Delta_h^-} + \frac{\Delta_h^-}{\Delta_h^+} (1 - \bar y)^{\frac{3}{2} \Delta_h^+}  \right) \,,\label{eq:Gh_graviton}  
\end{eqnarray}
where we have defined
\begin{equation}
\Delta_h^\pm = \pm \delta_h - 1  \,, \qquad   \delta_h = \sqrt{1-(4/9) \cdot p^2/\eta^2 } \,. \label{eq:Delta_h}
\end{equation}
 In particular, in the limit $y \to y_0 = 0$, one finds
\begin{eqnarray}
G_h(y_0,y^\prime) &=& -\frac{2}{3} \frac{1}{\eta} \frac{1}{\Delta_h^+} (1 -\bar y^\prime)^{\frac{3}{2}\Delta_h^+} \,. \label{eq:Ghz0zp}
\end{eqnarray}

We can consider, in particular, the analytical expressions for the IR-to-IR Green's functions, useful when the SM is localized on the IR brane, as is the case in the present model. These are given by
\begin{eqnarray}
%G_h(y_0,y_0;p) &=& - \frac{2}{3\rho}  \frac{1}{\Delta_h^+}  \,, \\
%G_h(y_0,y_1;p) &=& -\frac{2}{3\rho}  \frac{1}{\Delta_h^+} e^{-\frac{3A_1}{2} \Delta_h^+}  \,,  \label{eq:Gh_z0z1_asymp}  \\
G_h(y_1,y_1;p) &=&  \frac{1}{ 3 \eta}  \frac{e^{3A_1
}}{\delta_h} \left( -1 + \frac{\Delta_h^-}{\Delta_h^+} e^{-3A_1\delta_h}  \right) \,. \label{eq:Gh_z1z1_asymp} 
\end{eqnarray}
All Green's functions include the zero-mode contributions, which behave as
\begin{equation}
G_h^0  = \lim_{p \to 0} G_h(y,y^\prime;p)=  \frac{3\eta}{p^2} \,,  \label{eq:Gh0}
\end{equation}
so that we can define the Green functions contributed to only by the continuum of KK modes, with the zero-mode contribution subtracted out, as
\begin{equation}
\mathcal G_h(y,y')\equiv G_h(y,y')-G_h^0 \,.
\label{eq:Gnozeromode}
\end{equation}

\subsection{\it Coupling of the Graviton with Matter Fields}
\label{subsec:coupling_to_matter}

We are assuming fields located in the brane $y=y_a$. Then, the usual form of the interaction Lagrangian in the 4D effective theory is given by the Lagrangian
\begin{equation}
\mathcal L_{5D} = \frac{1}{M_5^{3/2}} T^{\mu\nu}(x,y) h_{\mu\nu}(x,y) \delta(y-y_a)\,,
\label{eq:coupling}
\end{equation}
where $T_{\mu\nu}(x,y_a)$ is the energy-momentum tensor of the matter fields localized at $y_a$~\endnote{We are assuming here the simplified case where matter lives in some brane, as, e.g.,~the SM that is living in the IR brane, or perhaps some dark sector that could live in the UV brane. For matter (SM singlets) propagating in the extra dimension, one should replace the interaction term in Equation~(\ref{eq:coupling}) by $\int_0^{y_s} dy \, T^{\mu\nu}(x,y)h_{\mu\nu}(x,y)$.}. 

In particular, for the SM fields living in the IR brane at $y=y_1$, the energy-momentum tensor is given by
\begin{equation}
T_{\mu\nu}=D_\mu H^\dagger D_\nu H+i\bar\psi \gamma_\mu D_\nu\psi - F_\mu{}^\rho F_{\nu\rho}-\eta_{\mu\nu}\mathcal L_{\rm SM} \,,
\label{eq:Tmunu}
\end{equation}
where $D_\mu$ is the SM covariant derivative, $\psi$ corresponds to all SM left- and right-handed fermions, and $F_{\mu\nu}$ is the field strength of different gauge fields $F_{\mu\nu}=W_{\mu\nu}^a,B_{\mu\nu}$. The term proportional to $\eta_{\mu\nu}$ does not contribute to the different vertices, as the tensor $h_{\mu\nu}$ is traceless. Notice that, in the broken phase, when $\langle H\rangle = ( v/\sqrt{2} )  (0,1)^T$, the massive gauge bosons make contributions to $T_{\mu\nu}$ proportional to $m_V^2 V_\mu V_\nu$.
The graviton zero-mode wave function, $h^0_{\mu\nu}(x,y)= h_0(y)h^0_{\mu\nu}$, where $ h_0(y)=\sqrt{3\eta}$ is canonically normalized as in Equation~(\ref{eq:grav_norm}), couples with the energy-momentum tensor at the IR brane as
\begin{equation}
\frac{1}{M_{\rm Pl}} T^{\mu\nu}(x,y_1)h_{\mu\nu}^0(x)\,.
\end{equation}

We will now consider the coupling with matter of the continuum of KK modes, with Green's function $\mathcal{G}_h(y,y')$. The effective field theory (EFT) for matter localized at the brane~$y_a$, for momenta $p\ll \eta$, provides the dimension-eight operator $\mathcal O_h(x,y_a)$ with the Wilson coefficient $c(y_a)$ as
\begin{align}
\mathcal L_{\rm EFT}(y_a)&=c(y_a) \mathcal O_h(x,y_a) \,,\nonumber\\
 \mathcal O_h(x,y_a)&=T^{\mu\nu}(x,y_a)D_{\mu\nu,\rho\sigma}T^{\rho\sigma}(x,y_a)=T^{\mu}_{\ \nu} T_{\ \mu}^{\nu} -\frac{1}{3} (T^{\mu}_{\ \mu})^2 \,,
 \label{eq:LEFT}
\end{align}
where the Wilson coefficients here have a mass dimension of $-4$. 

In particular, for the SM that is localized in the IR brane $y=y_1$,
\begin{equation}
c(y_1)=-\frac{1}{6} \frac{1}{\eta^4} \,.
\end{equation}
Thus, gravitational interactions are suppressed by the TeV scale $\eta$, reflecting the fact that the continuum of KK modes is localized toward the IR. On the contrary, if there is some extra matter localized in the UV brane, as, e.g., SM singlets of a dark sector, the corresponding Wilson coefficient would be
\begin{equation}
c(y_0)=-\frac{1}{9}\frac{1}{\eta^2 M_{4}^2}\,.
\end{equation}
In fact, if we define the effective coupling $g_{\rm eff}(y_a)$ as 
\begin{equation}
|c(y_a)|\equiv g_{\rm eff}^2(y_a)\frac{1}{\eta^2} \,,
\end{equation}
we can see that $g_{\rm eff}(y_0)\simeq 1/M_{4}$, while $g_{\rm eff}(y_1)\simeq 1/\eta$.

\subsection{Low Energy Constraints}
The effective Lagrangian in Equation~(\ref{eq:LEFT}) does give rise, in particular, to the dimension-eight operators, as in the notation of Refs.~\cite{Eboli:2006wa,Almeida:2020ylr},
\begin{equation}
\mathcal L_{\rm EFT}\supset \sum_{i=0}^2\frac{f_{S_i}}{\eta^4}\mathcal O_{S_i} \,,
\end{equation}
where
\begin{equation}
\begin{aligned}
  \mathcal O_{S_0} &= (D^\mu H^\dagger D^\nu H)(D_\mu H^\dagger D_\nu H) \,,\quad \\
  \mathcal O_{S_1} &= (D^\mu H^\dagger D_\mu H)(D^\nu H^\dagger D_\nu H) \,,  \\
\mathcal O_{S_2} &= (D^\mu H^\dagger D^\nu H)(D_\nu H^\dagger D_\mu H)\,, \\
f_{S_0} &= -\frac{1}{12},\quad f_{S_1}=\frac{1}{18}\,,\quad f_{S_2}=-\frac{1}{12}\,.
\label{eq:operatorsS}
\end{aligned}
\end{equation}
The contributions of these effective operators to the observables $S,T,U$ were computed in Ref.~\cite{Eboli:2006wa} as%~\endnote{We thank Prof.~O.~J.~P.~\' Eboli for a private communication on this result.}
\begin{equation}
\alpha T=-\frac{15}{16\pi^2}(m_W/\eta)^4\left(f_{S_0}+f_{S_2}+\frac{2}{5}f_{S_1}\right)(1+c_W^2)\frac{s_W^2}{c_W^2}\log(\eta/m_W) \,,
\end{equation}
where $\alpha$ is the fine structure constant, and $s_W (c_W)$ is the sine (cosine) of the electroweak mixing angle $\theta_W$, while $S=U=0$. Now, using the values in Equation~(\ref{eq:operatorsS}), we get
\begin{equation}
\alpha T\simeq \frac{13}{96 \pi^2}\left(m_W/\eta \right)^4(1+c^2_W)\frac{s^2_W}{c^2_W}\log (\eta/m_W) \,.
\end{equation}
From the experimental bound $T<0.054$, see Ref.~\cite{ParticleDataGroup:2024cfk}, one gets the mild bound $\eta\gtrsim 140$ GeV. The small value of the $T$ parameter occurs mainly because this effect stems from a dimension-eight operator, and is, thus, suppressed by the fourth power of $1/\eta$.

\subsection{High-Energy Constraints}

The effective Lagrangian in Equation~(\ref{eq:LEFT}) also gives rise to a number of dimension-eight operators, which contribute to an anomalous quartic gauge coupling (aQGC) as
\begin{equation}
\mathcal L_{\rm EFT}\supset \sum_j \frac{f_{T_j}}{\eta^4}\mathcal O_{T_j}+\sum_k \frac{f_{M_k}}{\eta^4}\mathcal O_{M_k} \,,
\end{equation}
where, using the notation of Refs.~\cite{Eboli:2006wa,Almeida:2020ylr}, we have
\begin{equation}
\begin{aligned}
\mathcal O_{T_0}&=(W^{\mu\nu}W_{\mu\nu})(W_{\alpha\beta}W^{\alpha\beta}) \,, \qquad \mathcal O_{T_2}=(W^{\mu\alpha}W_{\nu\alpha})(W_{\mu\beta}W^{\nu\beta}) \,,\\
\mathcal O_{T_5}&=(W^{\mu\nu}W_{\mu\nu})(B_{\alpha\beta}B^{\alpha\beta}) \,, \qquad \hspace{0.22cm} \mathcal O_{T_7}=(W^{\mu\alpha}W_{\nu\alpha})(B_{\mu\beta}B^{\nu\beta}) \,, \\
\mathcal O_{T_8}&=(B^{\mu\nu}B_{\mu\nu})(B_{\alpha\beta}B^{\alpha\beta}) \,, \qquad \hspace{0.43cm} \mathcal O_{T_9}=(B^{\mu\alpha}B_{\nu\alpha})(B_{\mu\beta}B^{\nu\beta}) \,,
\label{eq:operatorsT}
\end{aligned}
\end{equation}
and
\begin{equation}
\begin{aligned}
&\mathcal O_{M_0} = (W^{\mu\nu}W_{\mu\nu})(D^\alpha H^\dagger D_\alpha H)\,, \qquad \mathcal O_{M_1}=(W^{\mu\alpha}W_{\nu\alpha})(D_\mu H^\dagger D^\nu H) \,, \\
&\mathcal O_{M_2} = (B^{\mu\nu}B_{\mu\nu})(D^\alpha H^\dagger D_\alpha H) \,, \qquad \hspace{0.22cm}\mathcal O_{M_3} = (B^{\mu\alpha}B_{\nu\alpha})(D_\mu H^\dagger D^\nu H) \,,
\label{eq:operatorsM}
\end{aligned}
\end{equation}
with Wilson coefficients
\begin{equation}
\begin{aligned}
&f_{T_0}=\frac{1}{18} \,, \quad f_{T_2}=-\frac{1}{6} \,,\quad f_{T_5}=\frac{1}{9} \,,\quad f_{T_7}=-\frac{1}{3} \,,\quad f_{T_8}=\frac{1}{18} \,,\quad f_{T_9}=-\frac{1}{6} \,, \\
&f_{M_0}=-\frac{1}{9}\,, \quad f_{M_1}=\frac{1}{3} \,,\quad f_{M_2}=-\frac{1}{9} \,,\quad f_{M_3}=\frac{1}{3} \,.
\end{aligned}
\end{equation}

The LHC constraints on the above operators are obtained from the CMS experiment~\cite{CMS:2019qfk,CMS:2020ioi,CMS:2020gfh}. The strongest constraints are on the operators $\mathcal O_{T_0}$ and $\mathcal O_{T_2}$, which translate into the 95\% CL lower bound $m_g\gtrsim 1.3$ TeV. Projections in FCC-hh at $\sqrt{s}=100$ TeV and integrated luminosities of up to 30 ab$^{-1}$ have been made on the anomalous $WW\gamma\gamma$ couplings~\cite{Gutierrez-Rodriguez:2021hip}, which, for leptonic decay channels of the $W$s in the final state, yield future bounds reaching values of $m_g\gtrsim 7$ TeV.

Violation of unitarity is an inevitable consequence of the presence of aQGC, e.g.,~in longitudinal gauge boson scattering processes involving four vector particles, as the corresponding scattering amplitudes grow with $\hat s^2$, where $\sqrt{\hat s}$ is the center-of-mass energy, since the SM cancellation fails. Systematic analyses concerning this behavior for the operators $\mathcal O_{S_i}$, $\mathcal O_{T_j}$ and $\mathcal O_{M_k}$ have been thoroughly investigated in Refs.~\cite{Almeida:2020ylr,Guo:2020lim}. The unitarity violation indicates a failure of the EFT to describe the corresponding processes at such large values of $\sqrt{\hat s}$. In particular, using the general results in Ref.~\cite{Almeida:2020ylr}, the unitarity constraints imply an upper bound as $\sqrt{\hat s}\lesssim 2\, m_g$ for the validity of the EFT.

\section{Braneworld Cosmology}
\label{sec:braneworld}

In this section, we will consider a 5D model with just one brane where the SM is located, the UV brane, and the singularity. In this case, the graviton interactions with the SM are so feeble that the bulk system (the graviton and radion KK modes) at temperature $T_b$ is never in equilibrium with the SM plasma (which keeps its own temperature $T\neq T_b$). The bulk system will be dubbed the dark sector (DS). After time compactification, the bulk system is in equilibrium with the BH, such that $T_h=T_b$, and there is no conical singularity contribution to the free energy from the black hole. 

After cosmological inflation by the inflaton field, the SM plasma is heated by the inflaton interactions to the reheating temperature $T_R$, while the bulk is not (provided the inflaton only interacts with the SM), so its temperature is much lower than the SM one $(T_b \ll T_R)$. An energy density for the dark sector can only be provided by freeze-in, as we will see.

Here, it is convenient to use cosmological coordinates $r$ related to the proper coordinates~by
\begin{equation}
1-\frac{y}{y_s}=\left(\frac{r}{r_b} \right)^{\nu^2} \,, \qquad  \eta=\frac{1}{\nu^2 y_s} \,,
\end{equation}
where $r_b$ is the location of the brane, which corresponds to the UV brane $y=0$. Then, the metric can be written as
\begin{equation}
ds^2= g_{MN} dx^M dx^N=e^{-2A(r)}(h(r)d\tau^2 - d\vec{x}^2) - \frac{e^{-2B(r)}}{h(r)}dr^2   \,,
\end{equation}
with
\begin{align}
A(r)&=-\log(r/L)  \,,  \nonumber\\
B(r)&=-\nu^2\log(r/r_b)+\log(\eta r) \,, \quad \eta = k \, e^{\nu\bar v_b} \,, \\
h(r)&=1-\left(\frac{r_h}{r} \right)^{4-\nu^2} \,,\qquad  \bar\phi(r)=\bar v_b-\nu \log(r/r_b) \,,\nonumber
\end{align}
with $v_b$ being the value of the field $\phi$ at the brane, as fixed by the brane-localized potential. The $\eta$ parameter is a physical scale, {and $L$ is an integration constant with a dimension of length that can be fixed to any value.} The $h(r)$ function describes a planar black hole horizon located  at $r=r_h\leq r_b$ in the bulk. See \cite{Fichet:2023xbu} for a careful treatment of the gauge redundancies and integration constants. Notice that, with respect to the metric (\ref{eq:sol_A_T0}), we have introduced a non-vanishing constant value at the origin $y=0$: $A(r_b)$. With this choice, $a_b\equiv r_b/L$ will be identified with the scale factor when solving the equations of motion for the Friedmann--Lema\^{\i}tre--Robertson--Walker cosmology in the brane at $r=r_b$.

We can now apply the thermodynamic relations of Section~\ref{sec:thermodynamics} to cosmological coordinates, which yield the relations
\begin{align}
T_h&=\frac{4-\nu^2}{4\pi}\eta \left(\frac{r_h}{r_b} \right)^{1-\nu^2} \,, \label{eq:fromthermodynamics_Th} \\
s_{h}&=\frac{4\pi}{\kappa^2}\left(\frac{r_h}{r_b} \right)^{3} \,, \label{eq:fromthermodynamics_sh} \\
\rho_{h}&=\frac{3}{\kappa^2}\eta \left(\frac{r_h}{r_b} \right)^{4-\nu^2} \,, \label{eq:fromthermodynamics_rhoh}  \\
f_{h}&=  - P_h =-\frac{1-\nu^2}{\kappa^2}\eta \left(\frac{r_h}{r_b} \right)^{4-\nu^2} \,.  \label{eq:fromthermodynamics_fh}
\end{align}

Using this result, we can infer important information about the thermodynamic stability of the fluid. 
We can see that the free energy is negative (positive) for $\nu<1$ ($\nu>1$). 
This implies that the existence of the fluid, i.e.,~of the bulk black hole, is thermodynamically favored for $\nu<1$, while it is unfavored for $\nu>1$. Notice, however, that this conclusion assumes $r_h\ll r_b$. In the $\nu>1$ case, it turns out that the presence of the fluid becomes favored at larger values of $r_h$ \cite{Barbosa:2024pyn}.  

Finally, for $\nu=1$, 
\begin{align}
T_h&=\frac{3}{4\pi}\eta \,, \\
s_h&=\frac{4\pi}{\kappa^2}\left(\frac{r_h}{r_b} \right)^{3} \,, \\
\rho_h&=\frac{3}{\kappa^2}\eta \left(\frac{r_h}{r_b} \right)^{3} \,,  \label{eq:rho_h} \\
f_h&= - P_h =0 \,,
\end{align}
so that the free energy vanishes at leading order. Therefore,
the next-to-leading order term must be determined to assess stability, as conducted in Ref.~\cite{Fichet:2026nct}. 

\subsection{Braneworld Cosmology and Friedmann Equations}
\label{subsec:Friedmann_eq}

%From here on we will only consider the case of the linear dilaton background, $\nu=1$.
We assume that all the fields of the Standard Model are localized on the brane at $r=r_b$. The brane  $T^b_{\mu\nu}$ stress tensor is identified as the SM stress tensor, $T^b_{\mu\nu} \equiv T^{\rm SM}_{\mu\nu} $. Following Ref.~\cite{Fichet:2023xbu}, the effective Einstein equation has the form 
\begin{equation}
G_{\mu\nu} = \frac{1}{M_{4}^{2}} \left( T_{\mu\nu}^{\rm SM} + T_{\mu\nu}^{h}\right) + O\left( \frac{T_{\rm SM}^2}{M_5^{6}} \right) \,, \label{eq:D_1_Einstein}
\end{equation}
where the total stress tensor $T^{\textrm{tot}}_{\mu\nu} = T^{\rm SM}_{\mu\nu}+ T^{h}_{\mu\nu}$ contains the contribution $T^h_{\mu\nu}$ corresponding to the bulk stress tensor projected on the brane~\cite{Shiromizu:1999wj,Fichet:2022xol,Fichet:2022ixi}. The projection procedure leads to the structure of a 4D perfect fluid given by
  \begin{equation}
  T_\nu^{h\,, \mu} = g^{\mu\lambda} T^h_{\lambda\nu} = \textrm{diag}(-\rho_h,P_h,P_h,P_h) \,,
  \end{equation}
so that we will refer to $T^h_{\mu\nu}$ as the holographic fluid contribution to the stress tensor.

We assume that the brane is moving, so $r_b$ depends on time. Within the low-energy regime assumed in \eqref{eq:D_1_Einstein}, the brane motion is nonrelativistic, and simplifications occur; see \cite{Fichet:2022ixi, Fichet:2022xol} for details. 
 The effective Einstein equation, Equation~(\ref{eq:D_1_Einstein}), then produces the 
effective Friedmann equations  
\begin{equation}
3 M_4^2 H^2 = \rho_{\textrm{tot}} \,, \qquad  6 M_4^2 \frac{\ddot r_b(t)}{r_b(t)} = - \left( \rho_{\textrm{tot}} + 3 P_{\textrm{tot}}  \right) \,,
\end{equation}
where $r_b(t)/L\equiv a(t)$ plays the role of the scale factor from the induced metric, with $H = \dot r_b/r_b $  being the corresponding Hubble parameter. In these equations, we have $\rho_{\textrm{tot}} \simeq \rho_{\rm SM} + \rho_{h}$ and $P_{\textrm{tot}} =  P_{\rm SM} + P_h$. 

The energy density and pressure for the holographic fluid $(\rho_h,P_h)$ receive the following contributions
    \begin{align}
 \rho_h^W &= 3 P_h^W =   3 \left( 1 - \frac{\nu^2}{4}  \right)  \eta^2 M_4^2 \left( \frac{r_h}{r_b} \right)^{4 - \nu^2} \,, \\
 \rho_h^\phi &= - P_h^\phi =  -3 \eta^2 M_4^2 \times  \left[ 1 - \frac{\nu^2}{4} \left( \frac{r_h}{r_b} \right)^{4 - \nu^2}  \right]  \,, \\
 \rho_h^\Lambda &= -P_h^\Lambda = \frac{\Lambda_b^2 M_4^2}{12 M_5^6}  \,,
 \end{align}
corresponding to the projection on the brane of: (i)~the 5D Weyl tensor; (ii)~the bulk stress tensor; and (iii)~the brane tension, respectively. The summation of these contributions can be split into a vacuum (constant) and a fluid contribution, i.e.,
\begin{equation}
\rho_{h} = \rho_{\textrm{vacuum}} + \rho_{\textrm{fluid}}  \,, \qquad P_{h} = P_{\textrm{vacuum}}  + P_{\textrm{fluid}} \,.
\end{equation}
The fluid contributions turn out to be
\begin{equation}
\rho_{\textrm{fluid}} = 3 \eta^2 M_4^2 \left(\frac{r_h}{r_b}\right)^{4 - \nu^2} \,, \qquad P_{\textrm{fluid}} = (1-\nu^2) \eta^2 M_4^2 \left(\frac{r_h}{r_b}\right)^{4 - \nu^2}  \,,
\label{eq:fluid}
\end{equation}
while the vacuum contribution is given by
\begin{equation}
\rho_{\textrm{vacuum}} = - P_{\textrm{vacuum}} = \Lambda_4 M_4^2 \,,
\end{equation}
where the 4D cosmological constant $\Lambda_4$ and the Planck mass $M_4$ are related to the parameters of the bulk action as
\begin{equation}
\Lambda_4 =    - 3 \eta^2 + \frac{\Lambda_b^2}{12  M_5^6} \,, \qquad M_5^3 = \eta M_4^2 \sqrt{ 1 + \frac{\Lambda_4}{3\eta^2} } \,.
\end{equation}
The brane tension can be ultimately tuned to set $\Lambda_4$ to (almost) zero, so we will consider that the 5D Planck mass turns out to be related to the 4D version by $M_5^3 = \eta M_4^2$, a result that is consistent with Equation~(\ref{eq:Mpl}). Notice that these quantities are consistent with those obtained from the BH thermodynamics, Equations~(\ref{eq:fromthermodynamics_Th})--(\ref{eq:fromthermodynamics_fh}).

On the other hand, the SM fields form a thermal bath with temperature $T$; hence, $\rho_{\rm SM}=\rho_{\rm SM}(T)$ and $P_{\rm SM}=P_{\rm SM}(T)$. The brane position today (i.e.,~at $t=t_0$) is denoted by $r_b(t_0)=r_{b,0}$.\,\endnote{One could assume $a(t_0)=1$ without loss of generality, in which case, $r_b(t_0)=L$. 
We do not make this assumption in the present~work. }
Assuming adiabatic expansion of the universe, we have  $\frac{r_b(t)}{r_{b,0}} = \frac{ T_{0} }{T(t)}$,  where $T_{0} = 0.23 $~meV is
the temperature of the universe today.

\section{Dark Sector as a Holographic Fluid for the LD Background}
\label{sec:holo_fluid_DM}

In  the  background with $\nu=1$, the pressure of the holographic fluid vanishes up to %${\cal O}(r_h^3/r_b^3) $
${\cal O}(r_h^6/r_b^6) $  corrections (see Ref.~\cite{Fichet:2026nct}).
In the context of a cosmological braneworld model, we propose the identification\,\cite{Fichet:2023xbu,Fichet:2022xol} of the dark sector with a holographic fluid, which can eventually share all properties of dark matter.

The constant  temperature of the DS is  a hallmark of Hagedorn behavior. 
We remind the reader that the above thermodynamical properties match those of the thermal state of Little String Theory \cite{Aharony:1999ks,Kutasov:2001uf}.  In LST, the Hagedorn temperature is identified as  $T_H=\frac{M_s}{2\pi\sqrt{N}}$~\cite{Fichet:2023xbu}, where $M_s$ is the string scale.

Let us analyze the condition that guarantees that the holographic fluid can account for the dark matter abundance today. Using today's critical density $\rho_{c,0}= 3 H_0^2 M_4^2$, 
the DM abundance is given by  the remarkably simple formula
\begin{equation}
 \Omega_{\textrm{DM},0}=\left(\frac{\eta}{H_0} \right)^2\left(\frac{r_h}{r_{b,0}}\right)^3   \,. 
% \quad \quad  \rho_{\DM}(T_{\SM}) \simeq 3 \Omega_{\DM,0}H_0^2 M_4^2\, \frac{T_{\SM}^3}{T_{\SM,0}^3} \,.
\label{eq:rhoDMT}
\end{equation}
Fixing the dark matter abundance today to its observed value $\Omega_{\rm DM, 0}\simeq 0.27$ relates, therefore, the ratio $r_{h}/r_{b,0}$ to the $\eta$ parameter. Using today's value of the Hubble parameter $H_0\simeq 1.44\times 10^{-42}$ GeV, we obtain a bound on $r_h / r_{b,0}$. The bound can be extended to any time by considering the adiabatic expansion of the universe $(r_b \propto 1/T)$, from which we~obtain
\begin{equation}
\frac{r_h}{r_b} \simeq 3.5 \times 10^{-16} \frac{T}{\textrm{GeV}} \left(\frac{\textrm{GeV}}{\eta}\right)^{2/3}   \,.
\label{eq:rh_bound}
\end{equation}
As a result, the horizon is typically very far from the brane, and the
${\cal O}(r_h^3/r_b^3)$ corrections relative to the
  leading orders are tiny, even for the maximum reheating
  temperature $T_{R}$ that we find in Section~\ref{sec:freeze-in}.

Having identified the $\nu=1$ holographic fluid as  the cosmological dark matter observed  in our universe, we should  ask  about the cosmological history of such a scenario. Then, the following question arises: is there a mechanism capable of producing the holographic fluid with correct abundance in the early universe?  
In this section, we show that a mechanism analogous to freeze-in production~\cite{Hall:2009bx} naturally occurs in our holographic dark matter framework. 

Our main assumption is that the inflaton is brane-localized. As a result,  the brane position $r_b$ (i.e.,~the scale factor) grows exponentially during inflation, such that the bulk can be considered as empty at the end of inflation.  It is thus well-motivated to  consider the  system at the reheating time  with no bulk black hole, i.e.,~$r_h=0$. The corresponding reheating temperature is denoted by $T_{R}$.

\subsection{Heating the Bulk: Graviton Radiation from the Brane}
\label{se:graviton_emission}

The Standard Model particles on the brane couple to  bulk gravitons as
\begin{equation}
\mathcal L \supset \frac{1}{M_5^{3/2}}T^{\mu\nu}_{\rm SM}(x) h_{\mu\nu}(z_b,x)\,. 
\label{eq:5DLagrangian}
\end{equation}
Therefore, processes of the form $ \varphi + \bar\varphi \to h_{\mu\nu}$ (where $\varphi$ are SM fields) happen in the brane thermal bath. Since the gravitons live in the bulk, this is an emission mechanism through which energy is transferred from the brane to the bulk.

The rate of this energy transfer process, here denoted $\Delta \dot\rho_\SM$, was estimated in \cite{Gubser:1999vj} and computed in  AdS$_5$ in \cite{Hebecker:2001nv,Langlois:2002ke, Langlois:2003zb}.
% \,\endnote{We mention that an inaccurate high-mass approximation of the KK modes profiles was made in these references, with no qualitative consequences. The correct result can be  obtained by using a cut on  the full AdS$_5$ propagator, as done in App.\ref{app:gravitons}.    }
We performed a computation of $\Delta \dot\rho_\SM$ in the LD$_5$ background using a unitarity cut on  the LD propagators computed in \cite{Fichet:2023xbu, Fichet:2023dju}. The details of the computation are given in Ref.~\cite{Fichet:2026nct}. For center-of-mass energies much higher than the local 5D curvature $R|_{\rm brane} \approx -12 \eta^2$, the result matches the flat space result (as well as the AdS$_5$ with negligible curvature, as carried out in \cite{Hebecker:2001nv,Langlois:2002ke, Langlois:2003zb}). 

The brane-to-bulk energy transfer rate is 
\be
\Delta \dot\rho_\SM = -\frac{2\left(\sum_\varphi g_\varphi \kappa_\varphi^2 c_\varphi a_\varphi \right)}{5\pi^4 \eta M_4^2}\Gamma\left(\frac{7}{2}\right)\Gamma\left(\frac{9}{2}\right)\zeta\left(\frac{7}{2}\right) \zeta\left(\frac{9}{2}\right) T^8  \,, 
\label{eq:deltarho}
\ee
where $a_s=a_V=1$, $a_f\simeq 0.75$ ($s$, $V$ and $f$ stand for scalars, vectors and fermions, respectively), and $c_{s,f,V}=(1/12,1/16,1/4)$. Considering the SM degrees of freedom with $g_s=4$, $g_V=12$ and $g_f=45$, we get $\sum_\varphi g_\varphi \kappa^2_\varphi c_\varphi a_\varphi = 20.78$, where $\kappa_\varphi$ is the number of spin degrees of freedom of $\varphi$ ($\kappa_s=1$ and $\kappa_V=\kappa_f=2$). Numerically, 
\be
\Delta \dot\rho_\SM = - C_\rho \frac{T^8}{\eta M_4^2}\,,\quad \quad\quad  C_\rho \simeq 3.919 \,.
\label{eq:c}
\ee

\subsection{Holographic Dark Matter Freeze-In}
\label{sec:freeze-in}

The 4D effective Einstein equation, Equation \eqref{eq:D_1_Einstein}, leads to a 4D effective conservation equation. It can be derived either by contracting with $\nabla^\mu$ and using the 4D Bianchi equation or by starting with the 5D conservation equation and projecting it onto the brane; see \cite{Fichet:2022xol, Fichet:2022ixi} for details and checks. 
The effective conservation equations for the SM and holographic energy densities are 
\begin{align}
\dot{\rho}_\SM &=-4 H \rho_\SM + \Delta \dot \rho_\SM \,,   \label{eq:rhob}\\
\dot{ \rho}_h &=-3 H \rho_h + \Delta \dot \rho_h \,, \quad\quad  \Delta \dot \rho_\SM + \Delta \dot \rho_h = 0  \,.
\label{eq:rhoDM}
\end{align}

For the holographic fluid $\rho_h \propto r^3_h$ (see Equation~(\ref{eq:rho_h})), the evolution Equation~\eqref{eq:rhoDM} can alternatively be thought of as an evolution equation for the  $r_h$ parameter. That is, the equation must describe the evolution of the black hole horizon in the bulk, providing a geometric picture of freeze-in. The evolution equation for $r_h$ is found to~be
\be 
9  M_4^2 \eta^2 \frac{\dot r_h r^2_h}{r^3_b} =  \Delta \dot \rho_{h}\,. 
\label{eq:rh_evol}
\ee 
In Ref.~\cite{Fichet:2026nct}, we showed that this result can be directly derived from the bulk viewpoint.

%\subsubsection{Initial Condition}

We assume that our holographic model is valid during inflation. 
We further assume that the inflaton field  is brane-localized. Geometrically, this implies that $\dot r_b \propto e^{Ht}$, i.e., the brane runs away exponentially towards positive $r$. As a  result of the  braneworld inflation, we have $r_h/r_b\sim 0$, i.e.,~the holographic fluid is nonexistent at the end of inflation. {In this section, we call $r_h(T)$ the value of the horizon as a function of the temperature, while its constant value at low temperatures (i.e.,~in standard cosmology) will be denoted as $\bar r_h$. } 
We do not specify the inflation mechanism, and we simply start at the reheating time, with the corresponding temperature $T_{R}$ .\,\endnote{Notice, however, that the scenario is well-motivated by, e.g., the trace anomaly driven inflation mechanism of~\cite{Hawking:2000bb}.}
The initial condition for the dark matter energy density~is 
\be
\rho_{h}(T_{R} ) =0 \,. 
\label{eq:IC}
\ee

%\subsection{Holographic Dark Matter Freeze-In}

We now solve the evolution equations, Equations \eqref{eq:rhob} and \eqref{eq:rhoDM}. Let us  first notice that, for low temperatures, the energy transfer terms  $\Delta \dot \rho_{\SM,h}$  are negligible when compared to the other terms in the conservation equations. Because of the initial condition \eqref{eq:IC}, we can 
 also assume that $\rho_\SM\gg \rho_h$ throughout the evolution of the universe, such that 
the Hubble parameter is controlled by the SM thermal bath, 
\be
H=\frac{1}{\sqrt{3}M_4}\sqrt{\rho_\SM +  \rho_h}\simeq \frac{1}{\sqrt{3}M_4}\sqrt{\rho_\SM}  \,. \label{eq:H}
\ee
These points imply that, at low temperatures, the  scaling of $\rho_\SM$ and $\rho_h$ from standard cosmology is recovered. We denote the corresponding quantities in standard cosmology with bars, 
\be
\bar \rho_\SM (T) \propto T^4\,, \quad \quad \bar \rho_h (T) \propto T^3   \,.
\ee
Furthermore, we have that $ |\Delta \dot \rho_{\rm SM}(T_R)| \ll H \rho_{\rm SM}(T_{R})$ at reheating. Therefore, the  correction  to the SM energy density is small, and we can write 
\be
 \rho_\SM (T)   \approx \bar \rho_\SM (T) = g_\star  \frac{\pi^2}{30}T^4
\ee
to a very good approximation.

Using these simplifications, we now solve the evolution Equation~\eqref{eq:rhoDM}. 
The result is 
\begin{equation}
\rho_{h}(T)  
  = \bar \rho_{h}(T) \times \left (1-\frac{T^3}{T_{R}^3}\right)\quad \textrm{with}\quad \bar \rho_{h}(T) = \frac{\sqrt{10} C_\rho }{\pi\eta M_4\sqrt{g_*}} T_{R}^3 \,  T^3 \,.
\label{eq:rho_DM}
\end{equation}
This result constitutes the prediction  for the dark matter energy density in our holographic model. 
From \eqref{eq:rho_DM}, we can see that the freeze-in production occurs within roughly one order of magnitude in temperature below $T_{R}$. 

\begin{figure}[H]
%\begin{center}
 \includegraphics[width=0.45\textwidth]{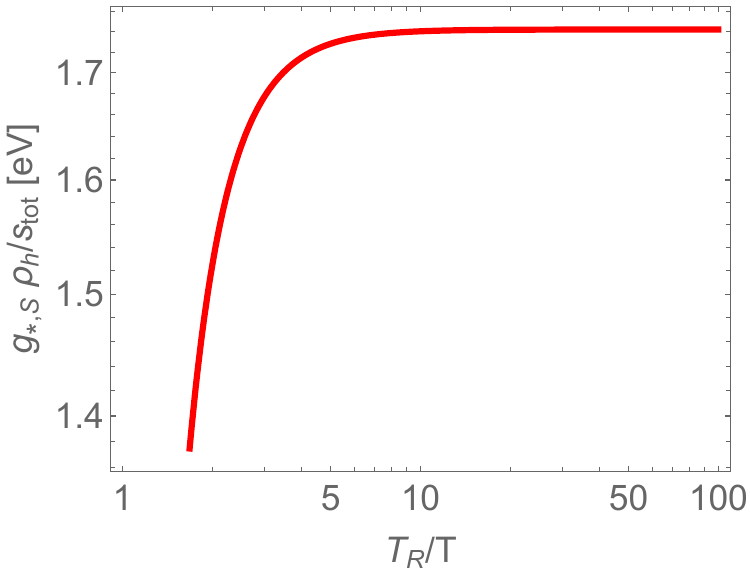}   \hspace{0.5cm}
 \includegraphics[width=0.455\textwidth]{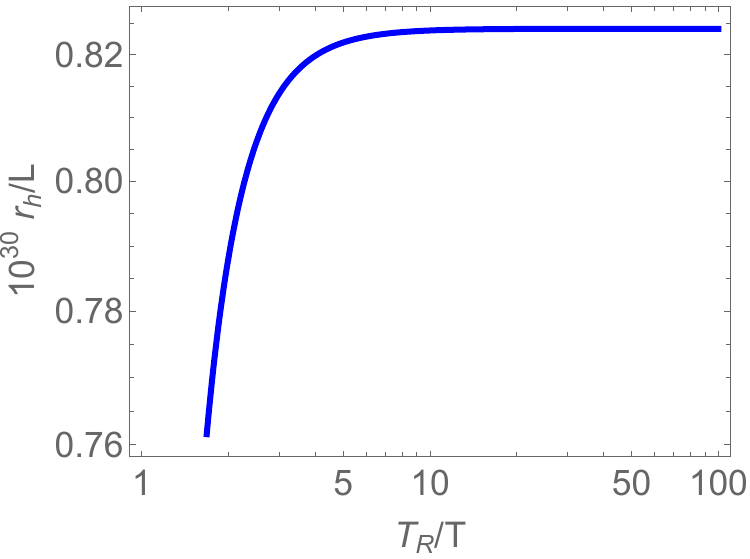}  
%\end{center}
     \caption{(\textbf{Left panel}): Energy-to-entropy ratio $\rho_{h} / s_{\textrm{tot}}$ in eV, normalized by $g_{\star,S}$, as a function of $T_{R} / T$. This quantity freezes at temperature $T\ll T_{R}$. (\textbf{Right panel}): Plot of $10^{30} r_h/L$ for $\eta=1$ TeV.
     }
     \label{fig:freezein}
 \end{figure}

Identifying with \eqref{eq:rhoDMT}, the dark matter abundance predicted by freeze-in is found to~be 
\be
\Omega_{{\rm DM},0} = \frac{\sqrt{10}C_\rho}{3 \pi  \sqrt{g_\star} \eta H_0^2} \left(\frac{T_{R}T_{0}}{M_4}\right)^3 \,. \label{eq:Omega_DM}
\ee
Finally, the energy-to-entropy density ratio is given by 
\be
\frac{\rho_{h}}{s_{\rm tot}}= \frac{45\sqrt{10}C_\rho}{2\pi^3 \sqrt{g_\star} \, g_{\star,S}} \frac{T^3_{R}-T^3}{ \eta M_4} \label{eq:ratio_result} \,,
\ee
where we used $s_{\rm tot}\approx s_{\SM} = g_{\star,S}\frac{2\pi^2}{45} T^3$, and  $g_{*,S}$ is the entropy effective number of degrees of freedom in the SM. 
This result is shown in the left panel of Figure~\ref{fig:freezein}. 

The freeze-in production of the dark fluid can equivalently  be interpreted as the freeze-in production of the black hole horizon, as given by (\ref{eq:rh_evol}), with solution
\be
r_h^3(T)=\bar r_h^3 \times \left(1-\frac{T^3}{T_{R}^3}\right)\,, \qquad \bar r_h^3= \frac{\sqrt{10}C_\rho}{3\pi \sqrt{g_*}}\left(\frac{L T_{0}}{\eta M_4} \right)^3 T_{R}^3\,,
\ee
from where we can see that $\bar r_h=r_h(T\ll T_{R})$. Then, (\ref{eq:rho_DM}) can be written as
\be
\rho_{h}(T)=3 \eta^2 M_4^2 \left(\frac{r_h(T)}{r_b(T)} \right)^3\,.
\ee
A plot of $r_h/L$ is exhibited in the right panel of Figure~\ref{fig:freezein}.

From \eqref{eq:Omega_DM}, we relate the reheating temperature to the model parameters. Keeping $M_5$ as the free parameter, we find 
\be
T_{R} = \left[ \frac{3 \pi \Omega_{\rm DM,0} \sqrt{g_*}}{C_\rho \sqrt{10}} \eta H_0^2 \right]^{1/3} \, \frac{M_4}{T_{0}} \simeq 9.4\times 10^{-10} M_5 \,.
\label{eq:TR}
\ee

Moreover,  considering the bound $M_5\lesssim M_4$ (which corresponds to $\eta\lesssim M_4$) yields, for the reheating temperature, the global upper bound 
\be T_{R} \lesssim 2.3\times 10^9~{\rm GeV}\,. 
\ee 
This is in good agreement with many realistic inflationary models~\cite{Cook:2015vqa}.

Finally, the value of the reheating temperature $T_{R}$ in (\ref{eq:TR}) needs to satisfy the consistency requirement $\rho_{\rm SM}(T_{R})\lesssim \eta^2 M_4^2$, 
 ensuring the validity of the low-energy approximation 
used  in the effective Einstein equation, Equation (\ref{eq:D_1_Einstein}). This translates into the mild (sufficient) condition $M_5\gtrsim 10$ GeV, valid for $T<T_{R}$. 

As expected for freeze-in production, the DM sector is never in thermal equilibrium with the SM, as condition (\ref{eq:TR}) guarantees that $\rho_{h}\ll \rho_{\rm SM}$ at temperatures $T\leq T_{R}$ (radiation domination). This also justifies one of the assumptions made in
the solving of the evolution equations. The remaining assumption, that $|\Delta\dot\rho_{\rm SM}|\ll H\rho_{\rm SM}$ for any $T\leq T_{R}$, is verified explicitly by using the relation \eqref{eq:TR} in the explicit expression for $\Delta\dot\rho_{h}$ given in \eqref{eq:c}.

\section{An Isolated Massive Graviton as Dark Matter}
\label{sec:isolated}

As we have seen, for the LD background, the graviton propagator is a gapped continuum, with a mass gap of $m_g=3\eta/2$. It has the same structure as an unparticle propagator~\cite{Georgi:2007ek,Georgi:2007si}, for which it was proven, after the introduction of radiative corrections into the self-energy, to be able to exhibit an isolated pole near (but below) the mass gap~\cite{Delgado:2008gj,Megias:2023kpk}. The main difference between the graviton and a general unparticle is that, for the former, the interactions with the SM fields are fixed by the single scale $M_5$ (i.e.,~$\eta$), and the dimension is also fixed, $d_\mathcal U=3/2$. In this section, we will propose that the isolated massive graviton that appears after the introduction of radiative corrections has the capabilities of being a feebly interacting dark matter candidate~\cite{Megias:2026qev}.

\subsection{The Graviton Propagator} 
\label{subsec:graviton_propagator}

We will now study the brane-to-brane graviton propagator in the LD model in the presence of a single brane at $r = r_b$, with the matter located at the brane. This follows similar steps to those in Section~\ref{sec:graviton}, so we omit most of the technical details.

%where the matter is located at the brane. %at $r=r_b$.
The system can also be described in conformal coordinates $z$, with solution%~\endnote{We denote the 4D comoving coordinates as $x^\mu$ and the physical ones as $\tilde x^{\mu}$. Note that the relation between comoving and physical coordinates is $
%\tilde x^\mu=a x^\mu,\; \tilde x_\mu=\frac{1}{a}x_\mu
%$,
%
%where we have used that comoving (physical) coordinates change with the metric $g_{\mu\nu}$  ($\eta_{\mu\nu}$). Therefore, while comoving coordinates with contravariant indices are conserved, comoving coordinates with covariant indices are not, as $x_\mu=a^2 \eta_{\mu\nu}x^\nu$. In addition, the condition $\tilde x^2=x^2$ is satisfied. Finally note that the fifth coordinate is also physical, $z=\tilde x^5$. %Similarly the conserved comoving momenta are $p_\mu$, with covariant indices. Comoving momenta with contravariant indices are not conserved as $p^\mu=\frac{1}{a^2}\eta^{\mu\nu}p_\nu$. Comoving momenta transform with the metric $g_{\mu\nu}$. Physical momenta are then $\tilde p_\mu=\frac{1}{a}p_\mu$ and $\tilde p^\mu=a p^\mu$ and transform with the metric $\eta_{\mu\nu}$. For the quadratic invariant they satisfy the condition $\tilde p^2=p^2$. Notice finally that comoving momenta are tied to the expanding coordinate grid, while physical momenta are what a local observer would actually measure with a detector.
%}
%
\begin{eqnarray}
  ds^2 &=& e^{-2A(z)} \left( \eta_{\mu\nu} d\tilde x^\mu d\tilde x^\nu  - dz^2 \right)  \,,  \label{eq:ds2z}\\
 A(z) &=& \eta (z - z_b)  \,, \qquad  \bar\phi(z) =  \bar v_b + \eta (z  - z_b)  \,, \qquad \eta = k \, e^{\bar v_b} \,,  \label{eq:ds2z2}
\end{eqnarray}
where $\tilde x^\mu = a(z_b) x^\mu$ are the 4D physical coordinates, with $a(z_b) = e^{-\eta z_b}$ being the scale factor. The relation with the brane cosmology coordinates is $\eta z=-\log(r/L)$. The brane position at $r=r_b$ then corresponds to $z_b= - \frac{1}{\eta} \log(r_b / L)$, while the singularity at $r_s=0$ corresponds to $z_s=\infty$. %In this way the space $\mathcal M^-$  corresponds to the interval $z\in[z_b,\infty)$, while the space $\mathcal M^+$ to the interval $z\in(-\infty,z_b]$.%~\endnote{In the presence of a black hole, the horizon is in the $\mathcal M^-$ space, at $z_h=- \frac{1}{\eta} \log(r_h/L)$, and the distance between the brane and the horizon behaves in terms of the temperature $T$ as $z_h(T) - z_b(T) \simeq - \frac{1}{\eta}\log(T_R T/(\eta M_4))$. Then, the horizon gets further and further away from  the brane as $T\to 0$.}

In terms of metric perturbations, the graviton is a transverse traceless fluctuation parametrized as
\begin{equation}
ds^2 = e^{-2A(z)} \left[ \left(\eta_{\mu\nu} + 2\kappa h_{\mu\nu}(\tilde x,z) \right) d\tilde x^\mu d\tilde x^\nu - dz^2  \right] \,, \label{eq:ds2_graviton_v2}
\end{equation}
where we are considering conformal coordinates. If one works with the rescaled field defined by $\tilde h_{\mu\nu}(\tilde x,z) = e^{-3A(z)/2} h_{\mu\nu}(\tilde x,z)$, the corresponding 5D graviton propagator %in the subspace $\mathcal M^\pm$ 
is given by
\be
G_{\mu\nu;\alpha\beta}(z,z^\prime;p^2) = G_h(z,z^\prime;p^2) P_{\mu\nu;\alpha\beta}(p^2) \,,
\ee
where $P_{\mu\nu;\alpha\beta}(p^2) $ is given by Equation~(\ref{eq:P}), 
%
%\begin{equation}
%P_{\mu\nu;\alpha\beta}(p^2) = \frac{1}{2}\left(\Theta_{\mu\alpha}\Theta_{\nu\beta}+\Theta_{\mu\beta}\Theta_{\nu\alpha} \right)-\frac{1}{3}\Theta_{\mu\nu}\Theta_{\alpha\beta},\quad \Theta_{\mu\nu}=\eta_{\mu\nu}-\frac{p_\mu p_\nu}{p^2}  \,,
%\end{equation}
while the coupling of the graviton $h_{\mu\nu}$ with the SM fields is driven by the conserved energy-momentum tensor $T^{\mu\nu}_{\rm SM}$, i.e.,
\be
\mathcal L_{\textrm{h-matter}}=\frac{1}{M_5^{3/2}}\, T^{\mu\nu}_{\rm SM}(\tilde x) h_{\mu\nu}(z_b,\tilde x) \,. \label{eq:couplinghT}
\ee

After solving the equations of motion of the graviton with Neumann boundary condition on the brane $\partial_z G(z,z^\prime)|_{z = z_b} = 0$, and with the regularity of the solution in the limit $z \to \pm \infty$, one finds that the scalar part of the 5D brane-to-brane graviton propagator is given by~\cite{Megias:2021mgj,Fichet:2026nct}
\be
G_{h}(z_b,z_b;s)= - \frac{ 1 }{m_g} \frac{1}{\Delta - 1} \,,
\ee
where
\begin{equation}
\Delta = \sqrt{1 - \frac{s}{m_g^2}} \,, \qquad \left(s \equiv p^2,\ m_g=\frac{3}{2}\eta\right) \,.
\end{equation}
There is a continuum of KK graviton modes for $m > m_g$, where $m_g $ is the mass gap. There is also a massless mode corresponding to the physical graviton in 4D. This can be seen by computing the limit of the propagator for $s\to 0$, for which one gets
\be
G_{h}(z_b,z_b;s) \stackrel[ s \to 0 ]{}{\simeq} \frac{2m_g }{s} \,.
\label{eq:graviton}
\ee

We will define the graviton propagator with the zero mode subtracted out
\begin{equation}
\mathcal G_h(z_b,z_b;s) \equiv G_h(z_b,z_b;s) - \frac{2 m_g}{s}  = -\frac{1}{m_g} \frac{1}{\Delta + 1} \,. \label{eq:Gh}
\end{equation}
In the following, we will consider the propagator given by Equation~(\ref{eq:Gh}) (which does not contain the massless graviton propagator), as the physical graviton mass is protected by diffeomorphism (general coordinate) invariance ($h_{\mu\nu}\to h_{\mu\nu}+\partial_\mu\xi_\nu+\partial_\nu \xi_\mu$), in much the same way as how gauge invariance protects the photon mass in QED.

%Finally notice that we have used physical coordinates in the propagator calculation, so the momenta which appear are physical ones, i.e.,~the momenta that would be measured by a local experiment. Still for computing physical processes, as cross-sections or interaction rates $\Gamma_{\rm int}=n\langle \sigma v\rangle$, which are much larger than the Hubble $\Gamma_{\rm int}\gg H$, we can consider the spacetime is Minkowski, as the characteristic timescale is $\tau_{\rm int}\simeq 1/\Gamma_{\rm int}\ll \tau_H\simeq 1/H$, and the processes take place essentially in flat spacetime. Of course, the expansion of the universe is properly taken into account in the Boltzmann equations when computing cosmological quantities as number, or energy, densities.

\subsection{The Isolated Resonance: The Massive Graviton}

 The brane-to-brane propagator of the massive gravitons was computed in Section~\ref{subsec:graviton_propagator} (see also \cite{Fichet:2026nct,Megias:2026qev}) and is given by
 \be
 G_h(s)=\frac{-1}{m_g+\sqrt{m_g^2-s}}  \,.
 \ee 
In this expression, for the propagator, any graviton zero mode has been subtracted out. 

Radiative corrections (see Ref.~\cite{Megias:2026qev}) generate an isolated pole in the second Riemann sheet of the complex $s$-plane at $s_p=m_p^2-i m_p \Gamma_p$ as the zeros of the equation
 \be
 D_h(s_p)-\Sigma_R(s_p)-i \Sigma_I(s_p)=0\,, \qquad  D_h(s)=-m_g-\sqrt{m_g^2-s}  \,,
 \ee
 with $\Sigma_R\equiv\textrm{Re} \, \Sigma$ and $\Sigma_I\equiv\textrm{Im} \, \Sigma$, where $\Sigma$ is the self-energy generated by the states localized in the brane, and $m_p<m_g$.
 
 We will consider the contribution of all states localized on the brane and simplify the analysis by assuming that the heaviest state contributing to the graviton self-energy has a mass of $m\gg m_g$. We will consider such a state as a heavy scalar $s=\phi$, e.g.,~the inflaton responsible for cosmological inflation (see Ref.~\cite{Megias:2026qev}). In that case, the full value of $\Sigma_R$ is dominated by this heaviest state, and, in fact, $\Sigma_R(s)\simeq\Sigma_R^s$ is a constant given by~\cite{Megias:2026qev}
 \be
 \Sigma_R^s \simeq -\frac{1}{32\pi^2}\left(5+4\log\frac{M_5^2}{m^2}  \right)\frac{m^4}{M_5^3}\,, \qquad \Sigma_I^s = 0  \,,
 \label{eq:SigmaRs}
 \ee
 where $m$ is the inflaton mass, while the lighter SM states coupled to the graviton can contribute to $\Sigma_I$.
 
 In Ref.~\cite{Megias:2026qev}, it was found that there is a pole in the second Riemann sheet of the complex $s$-plane parametrized by $a\in[1,2]$ at 
 \begin{equation}
 m_p=\sqrt{a(2-a)}\, m_g \,, \qquad
 \Gamma_p=-2\sqrt{\frac{(a-1)^2}{a(2-a)}}\Sigma_I(m_p^2)  \,,
 \end{equation}
  where the case $a=1$ corresponds to $m_p=m_g$, and $a=2$ to $m_p=0$, for a value of the inflaton mass $m$ given by
 %  condition (\ref{eq:condition2}) yields for the mass $m$ the value
 %
 \be
 \frac{m(a)}{M_5}=\left(\frac{e^{5/2}x}{-\mathcal W[-x]}  \right)^{1/4} \,, \qquad \textrm{with} \qquad x=\frac{16\pi^2 a\,m_g}{e^{5/2}M_5}  \,.
 \label{eq:maM5}
 \ee
 $\mathcal W$ is the principal branch of the Lambert function, so we can trade the parameter $m$ with~$a$. %We will postpone more details about the solution till the end of the section.

  To check the analytical approximation for the solution of the pole equation, we have considered a particular simple model where the real part $\Sigma_R$ is led by the heavy scalar $\phi$, so that we have already incorporated its effect on the propagator
  \begin{equation}
    \overline {G}_h(s) \simeq-\left(\sqrt{m_g^2-s}-(a-1)m_g+i\Sigma_I  \right)^{-1} \,. \label{eq:Gh_v2}
\end{equation}
 In this expression, $\overline G_h(s)$ stands for the full graviton propagator, including resummed loop contributions. The imaginary part of the self-energy is parametrized as
  \begin{equation}
    \Sigma_I(s)\equiv - \bar\gamma s^2/m_g^3 \,,
  \end{equation}
  where $\bar\gamma$ is an arbitrary dimensionless constant.

\begin{figure}[H]
%\begin{center}  
 \includegraphics[width=0.65\textwidth]{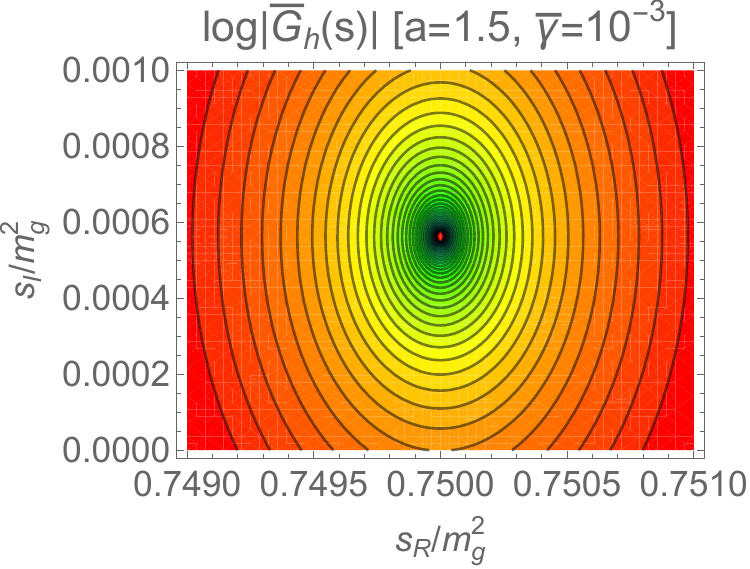}  %   \includegraphics[width=0.51\textwidth]{pole105}  
%\end{center}
     \caption{Contour lines of $|\overline G_h(s)|$ in the plane $s=s_R-i s_I$ for $a=1.5$ and $\bar \gamma =10^{-3}$.  At the pole: $s_R=m_p^2$ and $s_I=m_p\Gamma_p$. The red (green) regions indicate lower (higher) values of $|\overline G_h(s)|$.}
    \label{fig:pole}
 \end{figure}

  The
  pole mass and width are given by $m_p(a)=\sqrt{a(2-a)} \, m_g$, so
  that $\Sigma_I(m_p^2) = - a^2(2-a)^2\bar \gamma m_g$, and
  $\Gamma_p(a)=2a^{3/2}(2-a)^{3/2}(a-1) \bar\gamma m_g$. Contour lines
  of $|\overline G_h|$, where $\overline
  G_h=[G_h^{-1}(s) - \Sigma_R - i\Sigma_I(s)]^{-1}$, in the plane $s=s_R-i s_I$ are
  shown in Figure~\ref{fig:pole} 
  for the case where $a=1.5$ and $\bar\gamma =10^{-3}$. The analytical approximations of masses and widths are $m_p^2=0.75 m_g^2$ and $\Gamma_p\simeq 6.5\times 10^{-3} m_g$. These values are well reproduced by the pole shown in Figure~\ref{fig:pole}.

 The residue of the pole is given by
 \be
 R(m_p)=\frac{1}{D_h^\prime(s)-\Sigma^\prime_R(s)|_{m_p^2}}\simeq 2 (a-1)m_g \,,
 \ee
  and the coupling of the isolated resonance $\chi_{\mu\nu}$ to the energy-momentum tensor of matter in the brane is given by
  \be
  \kappa_\chi\equiv\frac{1}{M_5}\left(\frac{R(m_p)}{M_5} \right)^{1/2}\simeq \frac{\sqrt{3(a-1)}}{M_4} \equiv \frac{\lambda_\chi}{M_4} \,,
  \ee 
  where $\lambda_\chi$ is the Wilson coefficient, and the 4D Planck scale $M_4$ is the cutoff, defined in the interval $\lambda_\chi\in[0,\sqrt{3}]$.
  
  In fact, we can use the Wilson coefficient $\lambda_\chi$ as the free parameter, subject to the constraint $0<\lambda_\chi<\sqrt{3}$, and the pole mass and width of the isolated resonance are given~by
  \begin{align}
    m_p&=\sqrt{1-\frac{1}{9}\lambda_\chi^4}\,m_g \,, \qquad
    \Gamma_p=- 2\frac{\frac{\lambda_\chi^2}{3}}{\sqrt{1-\frac{1}{9}\lambda_\chi^4}}\,\Sigma_I(m_p^2)\,.
  \end{align}
  In particular, for $\lambda_\chi\ll 1$, we have that $m_p\simeq m_g$, while $\Gamma_p$ is strongly suppressed by $\sim \lambda_\chi^2$, a~property that will be used later on.
  
   \subsection{The Isolated Resonance as Dark Matter}
  
  Depending on the value of $\Gamma_p$, the resonance can decay on cosmological timescales or with a lifetime $\tau_p=1/\Gamma_p$ larger than the age of the universe. Only in the latter case can the resonance be a candidate for dark matter. In the former case, the resonance will decay and could eventually be detected at future accelerators or by its indirect effects. Both possibilities depend on the value of $\Gamma_p$, which, in turn, depends on the value of $\Sigma_I(m_p)$. In our model, the width is provided by the imaginary part of $\Sigma$ $(\Sigma_I)$, which is contributed to by the SM fields $\varphi$ (see Ref.~\cite{Megias:2026qev}) such that $m_p^2>4m_\varphi^2$.~\endnote{Note that every contribution corresponding to the SM field $\varphi$, with mass $m_\varphi$, contains a step function $\theta(s-4m_\varphi^2)$ and then vanishes for $s<4m_\varphi^2$.}
   
\begin{figure}[H]
%\begin{center}  
 \includegraphics[width=0.55\textwidth]{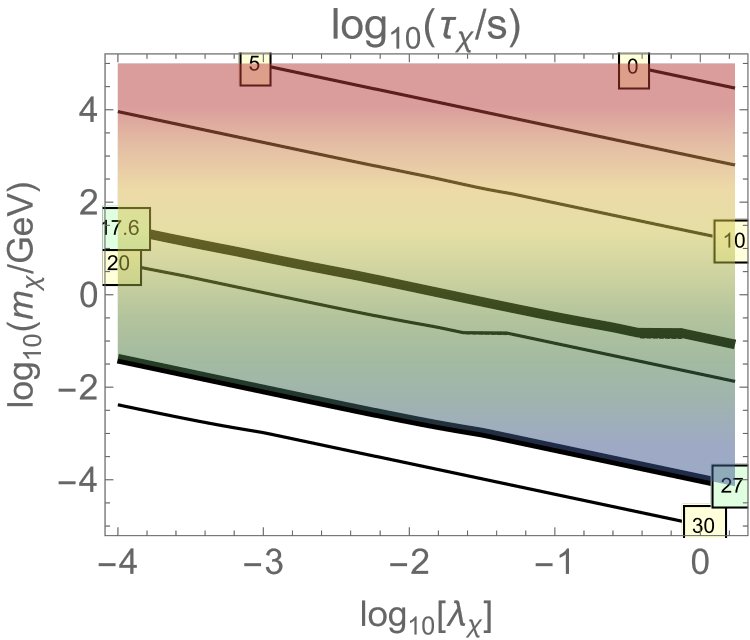}
%\end{center}
     \caption{Contour lines of  the lifetime $\tau_\chi$. The line for $\tau_\chi=10^{17.6}$ s corresponds to the present age of the universe, and $\tau_\chi=10^{27}$ s corresponds to the lower bound from cosmological observations. The red (blue) regions indicate lower (higher) values of $\tau_\chi$.}
     \label{fig:plottau}
 \end{figure}

   There are stringent bounds on decaying DM from indirect production and cosmological observations. In particular, DM can be constrained from observations of the galactic and extra-galactic diffuse X-ray and gamma-ray backgrounds, yielding upper bounds on $\tau_{\chi}$~\cite{Essig:2013goa}. Moreover, there are also cosmological constraints on exotic injection of electromagnetic energy and its effect on the CMB power spectra~\cite{Poulin:2016anj,Slatyer:2016qyl}, and effects of the decay channels $e^+ e^-$ and $\gamma\gamma$ in the 21 cm power spectrum~\cite{Facchinetti:2023slb,Sun:2023acy}. All of them fix a global  lower limit on the DM lifetime, which we take as $\tau_\chi\gtrsim 10^{27}$ s. In Figure~\ref{fig:plottau}, we plot contour lines of $\tau_\chi$ in the plane $(\lambda_\chi,m_\chi)$. The thick lines correspond to $\tau_\chi=\tau_U$, a lifetime equal to the universe lifetime, and $\tau_\chi=10^{27}~\textrm{s}$, the lower bound imposed by cosmological observations. Dark matter in the shaded region is then excluded.

  The isolated resonance $\chi_{\mu\nu}$ then corresponds to a feebly interacting massive particle (FIMP), a massive graviton that, in the unshaded region of Figure~\ref{fig:plottau}, has a lifetime much larger than the age of the universe, as $\tau_\chi>10^{27}$~s, and is thus a candidate for the dark matter of the universe. Assuming that the inflaton is mainly coupled to the SM fields (as it is only gravitationally coupled to the massive graviton), after inflation, at the reheating temperature $T_R$, the SM is a plasma in thermal equilibrium, while the FIMP is out of equilibrium with zero density. Its energy density must be produced by the freeze-in mechanism, as it happened with the holographic fluid.

  The process $\varphi\varphi\to\chi$ is too suppressed to yield a sizable energy density, given that the inverse process, the decay $\chi\to\varphi\varphi$, is required to be extremely small for the massive graviton not to decay on cosmological timescales. The freeze-in production by the 2-to-2 processes $\varphi_1\varphi_2\to\varphi_3\chi$ (where $\varphi_{1,2,3}$ are SM particles) was studied in great detail in Ref.~\cite{Cai:2021nmk}. It was found that the freeze-in is dominated by the channels involving the QCD coupling, in particular, $q\bar q\to g\chi$ and $qg\to q\chi$, where $g$ represents the gluons, and $q=c,b$ are heavy enough quarks to trigger the effect but light enough to be in thermal equilibrium below the critical temperature $T_c\simeq 160$ GeV of electroweak symmetry breaking~\cite{Quiros:1999jp}. Moreover, the resulting energy density is IR-dominated (the temperature integral is dominated by temperatures $T\sim T_c$) and, thus, insensitive to the reheating temperature (but sensitive to the value of the involved quark masses above $T_{\rm QCD}\simeq 150$ MeV), while the UV contribution is subleading. The final yield gives~\cite{Megias:2026qev}
  \be
  \Omega_\chi h^2\simeq 5.2\times 10^{-6}\lambda_\chi^2\, \left(\frac{\textrm{GeV}}{m_\chi}\right)^3  \,.
  \ee

\begin{figure}[H]
%\begin{center}  
  \includegraphics[width=0.55\textwidth]{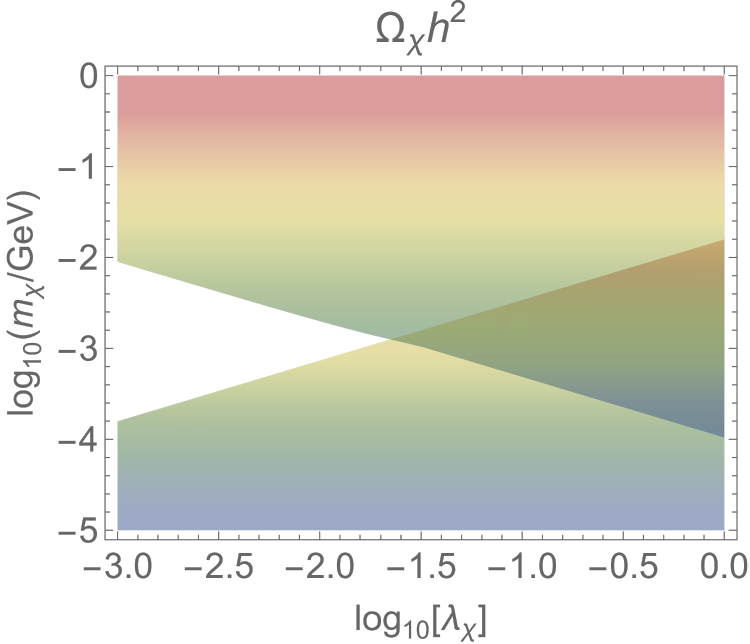}    % \includegraphics[width=0.45\textwidth]{plotOmegataump}
%\end{center}
     \caption{The white region in the plane $(\lambda_\chi,m_\chi)$ is allowed, as there, $\Omega_\chi h^2\leq 0.12$ and $\tau_\chi>10^{27}$~s. The red (blue) regions indicate lower (higher) values of either $\Omega_\chi h^2$ (lower shaded region) or $\tau_\chi$ (upper shaded region).}
     \label{fig:Omegachi}
 \end{figure}

  We show, in Figure~\ref{fig:Omegachi}, the available region in the plane $(\lambda_\chi,m_\chi)$. The lower shaded region is excluded by overclosure of the universe, as here, $\Omega_\chi h^2>0.12$, while the upper shaded region is excluded by cosmological observations, as here, $\tau_\chi<10^{27}$ s. 

The range for the isolated resonance, as well as the inflaton mass $m$, to be identified as DM is then
$
m_\chi\lesssim 2 \textrm{ MeV} \; \textrm{and} \;  m\lesssim 4\times 10^{11} \textrm{ GeV}\,,
$
while there is no theoretical lower bound, although Lyman-$\alpha$ forest data~\cite{Viel:2013fqw,Viel:2007mv}, as well as updated Milky Way satellite counts and strong gravitational lensing, put experimental lower bounds on warm dark matter (WDM) mass around $m_\chi\gtrsim 20$ keV~\cite{DEramo:2025jsb}, which corresponds to $m\gtrsim 8\times 10^{10}$ GeV. This yields the allowed bands in the fundamental parameters of our theory, which are summarized in Table~\ref{tab:summary}.
\begin{table}[H]
%  \centering
\caption{Summary of the range of values of the fundamental parameters of the theory.}
\label{tab:summary}
\begin{tabularx}{\textwidth}{LC}
    \toprule
    \multicolumn{2}{c}{Fundamental parameters} \\
  \midrule
     { DM mass }  &   $20\textrm{ keV}\lesssim m_\chi\lesssim 2\textrm{ MeV}$  \\
\midrule
        { 5D Planck mass }  &  $4 \times 10^{10} \textrm{ GeV} \lesssim M_5 \lesssim 2 \times 10^{11}\textrm{ GeV}$  \\
\midrule
 { Inflaton mass }    &  $8\times 10^{10} \textrm{ GeV}\lesssim m\lesssim 4\times 10^{11}\textrm{ GeV}$   \\
    \bottomrule
  \end{tabularx}

\end{table}

 \subsection{Braneworld Inflation}
 
 In the case of braneworld cosmology, where matter fields are confined to a three-brane in a warped 5D space,  the Hubble parameter in the brane was computed for the case of the linear dilaton background~\cite{Fichet:2022xol,Fichet:2023xbu,Fichet:2026nct} as %the Hubble parameter can be written as
\be
3H^2M_4^2=\rho_b\left(1+\frac{\rho_b}{2\lambda}\right)+\rho_{h}\,, \qquad \lambda=6 M_5^6/M_4^2 \,,\qquad \rho_{h}=\frac{\lambda}{2}(r_h/r_b)^3  \,,
\ee
where $\rho_b$ is the brane energy density, $r_h$ is the location of the horizon of the 5D black hole, and $\rho_{h}$, describes a holographic fluid that can play the role of holographic dark matter~\cite{Fichet:2026nct}. As this term scales as matter, its strength will be negligible in the early universe, during inflation, where we expect that $\rho_b \gg 2\lambda$, and we can approximate the energy density by using the inflaton $\phi$ potential $\rho_b\simeq V(\phi)$ so that 
\be
H^2\simeq \frac{V}{3M_4^2}\left(1+\frac{V}{2\lambda}  \right)\simeq \frac{V^2}{36 M_5^6}\,.
\label{eq:H-BC}
\ee

The slow-roll parameters for braneworld cosmology have been worked out in Refs.~\cite{Maartens:1999hf,Jaman:2018ucm}. For $V\gg 2\lambda$, they can be written as
\begin{align}
\epsilon(\phi)&\simeq 12 M_5^6 \frac{(V')^2}{V^3} \,, \\
\eta(\phi)&\simeq 12 M_5^6 \frac{V^{\prime\prime}}{V^2} \,, 
\end{align}
which yield the spectral tilt $n_s=1-6\epsilon+2\eta$. 

The end of inflation $\phi=\phi_f$ is fixed by the condition $\epsilon(\phi_f)=1$. The beginning of inflation $\phi=\phi_i$ is related to the number of e-folds before the end of inflation $N$ by
\be
N\simeq -\frac{1}{12 M_5^6}\int_{\phi_i}^{\phi_f}\frac{V^2}{V'}d\phi\,.
\ee
Finally, the CMB normalization of density perturbations yields
\be
A_s^2\simeq \frac{1}{(360\, \pi M_5^{9})^2}\frac{V^6(\phi_i)}{[V'(\phi_i)]^2} \simeq 2.1\times 10^{-9}  \,,
\ee
while the ratio of tensor-to-scalar perturbations~\cite{Jaman:2018ucm} is given by $r\simeq 24 \, \epsilon(\phi_i)$.

In Ref.~\cite{Megias:2026qev}, we studied, as an existence proof for the case of braneworld (non-standard) cosmology, a simple potential, given by~\endnote{In this section, we are just trying to provide the simplest inflationary potential in the brane that triggers the required properties of the isolated long-lived graviton dark matter. Finding a more motivated inflaton potential is outside the scope of the present paper and will be postponed to further studies.}
\be
V(\phi) = V_0 \times \left(\frac{\left(\frac{\phi}{\mu}\right)^2+\alpha}{\left(\frac{\phi}{\mu}\right)^2+1}-\alpha \right) \,, \label{eq:potential}
\ee
where $\alpha<1$ is a real parameter. 

In the UV regime, where $\phi/\mu\gg 1$, we get
\be
\epsilon\simeq \frac{96\mu^2 M_5^6}{m^2 \phi^6} \,,\qquad \eta\simeq -\frac{144 M_5^6}{m^2\phi^4} \,,\qquad N\simeq \frac{m^2}{192 M_5^6}\left(\phi_i^4-\phi_f^4 \right)  \,.
\ee
%
%where $\phi_i$ and $\phi_f$ denote the values of the inflaton at the beginning and the end of inflation, respectively, and $N$ is the number of e-folds spanned.
In the regime $1\ll \phi_f/\mu\ll\phi_i/\mu$, we~get
\be
\frac{\phi_f^6}{\mu^6}\simeq \frac{96 M_5^6}{m^2\mu^4} \,, \qquad \frac{\phi_i^4}{\mu^4}\simeq \frac{192 M_5^6}{m^2\mu^4}\,N  \,,
\ee
and the slow-roll parameters at the beginning of inflation $\phi_i$
\begin{equation}
\begin{aligned}
\epsilon_i&\simeq \frac{1}{16\sqrt{3}}\, \frac{m\mu^2}{M_5^3}\frac{1}{N^{3/2}}\quad \Rightarrow\quad r\simeq 24 \, \epsilon_i \simeq \frac{\sqrt{3}}{2}\, \frac{m\mu^2}{M_5^3}
\frac{1}{N^{3/2}} \,, \\
\eta_i&\simeq -\frac{3}{4N}\quad \Rightarrow\quad n_s\simeq 1-\frac{3}{2N} \,.
\end{aligned}
\end{equation}
We have to compare the prediction on the spectral index with the experimental values from Planck~\cite{Planck:2018jri}, $n_s=0.9649\pm 0.0042$ (leading to $N\simeq 43$), from ACT DR6~\cite{AtacamaCosmologyTelescope:2025blo}, $n_s=0.974\pm 0.003$ (leading to $N\simeq 58$), and from SPT-3G~\cite{SPT-3G:2025bzu}, $n_s=0.9679\pm 0.0033$ (leading to $N\simeq 47$), although some instability in the ACT DR6 and SPT-3G results has been recently reported in Ref.~\cite{Sabogal:2026qvy}.
Moreover, the value of $A_s^2$ is given~by
\be
A_s^2\simeq \frac{1}{1800\sqrt{3}\pi^2}\, \frac{m^5\mu^4}{M_5^9}N^{3/2}\simeq 2.1\times 10^{-9}  \,.
\label{eq:As2}
\ee

We can now particularize the previous results to our model, for which the inflaton mass is given in terms of $M_5$ as $m\simeq e^{5/8}M_5$, from Equation~(\ref{eq:maM5}). Then, the condition~(\ref{eq:As2})~gives
\be
\mu\simeq 8.8\times 10^{-3} M_5 \left( \frac{60}{N}\right)^{3/8}  \,,
\label{eq:muM5}
\ee
%
%thus evading Lyth's bound~\cite{Lyth:1996im},
and the values of $\phi_{i,f}$, for $N\simeq 60$, are given by
\be
\phi_f\simeq 0.36 M_5,\quad \phi_i\simeq 7.58 M_5  \,,
\ee
which satisfy the assumed condition $1 \ll \phi_{f}/\mu \ll \phi_{i}/\mu$ and, thus, evade the Lyth's bound~\cite{Lyth:1996im} on trans-Planckian inflaton excursion. Finally, the prediction for the tensor-to-scalar ratio is given by
\be
r\simeq 2.7\times 10^{-7}  \,.
\ee

Now, using the relation (\ref{eq:muM5}), we can easily check that the initial condition $V(\phi_i) \gg 2\lambda$ implies an upper bound on $M_5$ (and so on $m_g$), as%~\endnote{Use $M_5^3=\eta M_4^2=(2/3) m_g M_4^2$.}
\be
M_5\lesssim 3.4\times 10^{-3}M_4\simeq 8.2\times 10^{15}\textrm{ GeV} \,,
\ee
which is widely satisfied by our window in Table~\ref{tab:summary}. Furthermore, the Hubble parameter during inflation is given by
\be
  H_i = 2.3 \times 10^{-5} M_5 \lesssim 4.5\times 10^{6} \textrm{ GeV}\,,
\ee
where the last inequality comes from Table~\ref{tab:summary}. Similarly, the maximum value of the reheat temperature $T_R^{\rm max}$, assuming good reheating such that all the inflaton energy $V_i$ is converted into radiation, also gets the upper bound
\be
T_R^{\rm max}\lesssim  0.044 \, M_5\leq  8.9 \times 10^{9} \textrm{ GeV} \,,
\ee
%
%Notice that this maximum value of the reheating temperature is larger than the upper bound of Equation~(\ref{eq:boundonTR}), which is then permitted. 
which is satisfied by many realistic inflationary models~\cite{Cook:2015vqa}.

\section{Discussion}
\label{sec:discussion}

In this review, we have presented the capabilities of the linear dilaton 5D background from the point of view of particle physics and cosmology. 

Application to particle physics consists of models with two branes: the UV and the IR brane, where the Higgs is localized. This is an alternative to theories based on the AdS background. In proper coordinates, $y$, the theory extends from the origin $y=0$ to the singularity at $y_s\equiv 1/\eta$, which thus provides the fundamental scale of the theory $\eta$. For $\eta=\mathcal O$ (TeV), the 5D Planck scale is around $M_5\simeq 10^{-4} M_4$, and the warped factor solves the hierarchy between $\eta$ and $M_5$, while the string theory has to provide the solution to the hierarchy between $M_5$ and $M_4$.

The main application of the LD background to cosmology consists in describing unconventional, braneworld cosmology, with a single brane where matter is localized. Under exposure to the Friedmann equations, the location of the brane describes the scale factor of the universe under expansion. Thermodynamical quantities, or equivalently,
the Gauss-Codazzi equations and Israel junction conditions used to project the energy-momentum tensors on the brane, imply the presence of a pressureless holographic fluid that can play the role of dark matter. The difference with respect to the case of an AdS background, where the holographic fluid is dark radiation, is remarkable. Another remarkable property of the LD background is that the graviton propagator is a gapped continuum, unlike the AdS case, for which the graviton propagator is an ungapped continuum. Both facts {\it gapped vs. ungapped continuum } and {\it holographic dark matter vs. holographic dark radiation } are connected. In both cases, the dark energy density is zero after cosmological inflation and is then produced by UV freeze-in.

Moreover, for the case of the LD background, the presence of the mass gap in the propagator allows for the appearance of an isolated massive graviton, out of the continuum, via the resummation of self-energy radiative corrections. Its energy density is also zero after cosmological inflation, as it is feebly coupled to the SM and generated after the reheating temperature by freeze-in.
This provides a new (more conventional) candidate for dark matter, which we have proven to satisfy all theoretical, experimental and cosmological constraints in the window [20 keV, 2 MeV]. We have also found a surprising link between the possibility of the massive long-lived graviton as dark matter and the presence of an inflaton localized in the brane with a mass similar to the 5D Planck scale. We have presented, as a proof of existence, a simple model of inflation for the brane cosmology satisfying all cosmological observables and evading Lyth's bound.

Future research directions are clearly required to explore the properties and possible experimental detection of the candidates for dark matter we have presented: the holographic fluid and the long-lived isolated massive graviton. In fact, a very interesting possibility is that the holographic fluid and the isolated massive graviton might be independent components of dark matter such that $\Omega_h h^2+\Omega\chi h^2\simeq 0.12$, and we have found there is room in the parameter space where this possibility can be realized. The phenomenology of this composite dark matter is phenomenologically appealing and will be analyzed in the future. As for the experimental prospects, its direct detection seems out of the scope of present facilities, as they are feebly coupled to the SM. More promising are future measurements of deviations from Newton's law (fifth force experiments). On the other hand, before establishing the holographic fluid as a firm candidate for dark matter, the cosmological perturbation theory should be studied in detail. We leave it to future research.

\vspace{6pt}

 %%%%%%%%%%%%%%%%%%%%%%%%%%%%%%%%%%%%%%%%%%
\authorcontributions{Conceptualization, E.M. and M.Q.; methodology, E.M. and M.Q.; investigation, E.M. and M.Q.; writing---original draft preparation, E.M. and M.Q. All authors have read and agreed to the published version of the manuscript.}

\funding{The works of E.M. and M.Q. are supported by the ``Proyectos de
  Investigaci\'on Precompetitivos'' Program of the Plan Propio de
  Investigaci\'on of the University of Granada under grant
  PP2025PP-18. The research of M.Q. is also supported by the grant
  PID2023-146686NB-C31 funded by MICIU/AEI/10.13039/501100011033/ and
  by ERDF/EU, and under Severo Ochoa Centres of Excellence Programme
  2025-2029 (CEX2024001442-S). IFAE is partially funded by the CERCA
  program of the Generalitat de Catalunya.}

%\institutionalreview{Not applicable.}

%\informedconsent{Not applicable.}

\dataavailability{No new data were created or analyzed in this study. Data sharing is not applicable to this article.}

\acknowledgments{This review is dedicated to Ignatios Antoniadis on the occasion of his 70th birthday, in recognition of his many contributions to theoretical physics. One of the authors, M.Q., would like to express his gratitude for Ignatios' friendship, support, and guidance over so many years. We are also grateful for the valuable insights gained through discussions with Ignatios Antoniadis, Sergio Barbosa, Karim Benakli, Joan Cabrer, Marcela Carena, Leandro Da Rold, Antonio Delgado, Sylvain Fichet, Benjamin J. Galow, Gero von Gersdorff, Anish Ghoshal, Fotis Koutroulis, Germano Nardini, Jun Nian, Giuliano Panico, Ioannis Papadimitriou, Manuel P\'erez-Victoria, Hans-J\"urgen Pirner, Miguel Prieto, Stefan Pokorski, Oriol Pujol\`as, Rogerio Rosenfeld, Lindber Salas, Manuel Valle, Kambis Veschgini, Carlos Wagner, and Geovanna Yamanaki.}

\conflictsofinterest{The authors declare no conflicts of interest.} 

\appendixtitles{yes} % Leave argument "no" if all appendix headings stay EMPTY (then no dot is printed after "Appendix A"). If the appendix sections contain a heading then change the argument to "yes".
\appendixstart
\appendix
\section[\appendixname~\thesection]{Orbifold vs. Interval Picture}
\label{sec:interval-orbifold}
In this appendix, we will provide some simple ideas about the two possible (equivalent) descriptions in the case of $S^1/\mathbb Z_2$ compactifications. Considering the general metric 
\be
ds^2=e^{-2A(y)}\eta_{\mu\nu}dx^\mu dx^\nu-dy^2  \,,
\ee
there are two equivalent ways of describing the fifth dimension. We will do it with the simplest case of a bulk scalar field $\phi$, but the generalization to particles with spin is~straightforward.

\begin{description}
\item[{\it Orbifold picture:}] In this picture, we consider $y\in[-L,L]$ and identify the points $y\equiv -y$. There are then fixed points at $y=0,L$ (or equivalently at $y=0,-L$ as $-L\equiv L$). The physical space is then $S^1/\mathbb Z_2$. The field $\phi$ has a definite $\mathbb Z_2$-parity, either even [for which $\phi(-y)=\phi(y)$] or odd [for which $\phi(-y)=-\phi(y)$]. In the absence of brane terms in the action, this would automatically impose boundary conditions (BCs) at the fixed points, either Neumann $\partial_y\phi=0$ (for even fields) or Dirichlet $\phi=0$ (for odd~fields).

The KK decomposition in the orbifold picture
\be
\phi_{\rm orb}(x,y)=\sum_n \phi_n(x) f_{\rm orb}^{(n)}(y)
\ee
is such that the normalization of modes
\be
\int_{-L}^L dy \, e^{-2A(y)} f_{\rm orb}^{(n)}(y) f_{\rm orb}^{(m)}(y)=\delta_{mn}\ \Rightarrow \  2\int_{0}^L dy \, e^{-2A(y)} f_{\rm orb}^{(n)}(y) f_{\rm orb}^{(m)}(y)=\delta_{mn}
\label{eq:normalizationorbifold}
\ee
is satisfied, independently of the parity of the field.

\item[{\it Interval picture:}] In this picture, we consider $y\in[0,L]$, an interval with endpoints at $y=0,L$. The field $\phi$ has BC at the endpoints, either Neumann (if the field is even under $\mathbb Z_2$) or Dirichlet (if the field is odd under $\mathbb Z_2$), in the absence of brane terms. The KK decomposition in the interval picture
\be
\phi_{\rm int}(x,y)=\sum_n \phi_n(x) f_{\rm int}^{(n)}(y)
\ee
is such that the normalization of modes
\be
\int_{0}^L dy \, e^{-2A(y)} f_{\rm int}^{(n)}(y) f_{\rm int}^{(m)}(y)=\delta_{mn}
\label{eq:normalizationinterval}
\ee
is satisfied.
Comparison between (\ref{eq:normalizationorbifold}) and (\ref{eq:normalizationinterval}) gives then the relation
\be
f_{\rm orb}^{(n)}(y)=\frac{1}{\sqrt{2}}f_{\rm int}^{(n)}(y)\quad \textrm{or}\quad \phi_{\rm orb}(x,y)=\frac{1}{\sqrt{2}}\phi_{\rm int}(x,y)  \,.
\label{eq:relation}
\ee

\end{description}

The relation (\ref{eq:relation}) is easily translated to a similar relation for the Green's functions defined by the corresponding KK expansions
\be
G_{\rm orb}(p;y,y^\prime)=\sum_n\frac{f_{\rm orb}^{(n)}(y)f_{\rm orb}^{(n)}(y^\prime)}{p^2-m_n^2+i\epsilon}\quad\textrm{and}\quad G_{\rm int}(p;y,y^\prime)=\sum_n\frac{f_{\rm int}^{(n)}(y)f_{\rm int}^{(n)}(y^\prime)}{p^2-m_n^2+i\epsilon}\,.
\ee
Using relation (\ref{eq:relation}), we obtain
\be
G_{\rm orb}(p;y,y^\prime)=\frac{1}{2}G_{\rm int}(p;y,y^\prime)  \,.
\ee

This factor of 2 disappears in the physical amplitudes. For instance, if we couple the field $\phi$ to a source $J(x)$ localized at the brane $y=y_b$, the interaction can be written as
\be
\mathcal S_{\rm orb}=\int d^4x \, \lambda_{\rm orb} J(x) \phi_{\rm orb}(x,y_b) \qquad \textrm{and} \qquad \mathcal S_{\rm int}=\int d^4x \, \lambda_{\rm int} J(x) \phi_{\rm int}(x,y_b)  \,,
\ee
such that, when using Equation~(\ref{eq:relation}), we get the relation between the couplings
\be
\lambda_{\rm orb}=\sqrt{2}\lambda_{\rm int}  \,,
\ee
which guarantees that physical amplitudes are equal in both pictures.

The considered case of a scalar field $\phi$ can be easily generalized to fields with arbitrary spin as gauge bosons, fermions and the graviton.
\section[\appendixname~\thesection]{Solutions in the AdS-\boldmath{$\nu$} Model}
\label{sec:AAdS}
%\subsection[\appendixname~\thesubsection]{}

We will show in this appendix the solutions of the EoM in the asymptotically AdS $\nu$-model (AdS-$\nu$). This model is defined by the following superpotential and potential~\mbox{\cite{Cabrer:2009we,Megias:2019vdb,Fichet:2022ixi}}
\begin{equation}
W(\bar\phi) = \frac{6k}{\kappa^2}\left( 1 + e^{\nu\bar\phi} \right) \,,  \qquad V(\bar\phi) = -\frac{6k^2}{\kappa^2} \left[  1 + 2 e^{\nu \bar\phi}  + \frac{(4-\nu^2)}{4} e^{2\nu\bar\phi} \right] \,, \label{eq:W_AdS_nu}
\end{equation}
which behave as the corresponding versions for the AdS metric in the limit where $\bar\phi\to-\infty$.
The background solution in proper coordinates is given by
\begin{align}
A(y)&= k y -\frac{1}{\nu^2}\log\left(1-\frac{y}{y_s} \right) \,, \label{eq:sol_AdSnu_T0_A} \\
\bar\phi(y)&=-\frac{1}{\nu}\log[\nu^2 k(y_s-y)] \,. \label{eq:sol_AdSnu_T0_phi}
\end{align}
The value of the Ricci scalar is then given by
\begin{equation}
R=20k^2\left[ 1+2e^{\nu\bar\phi}+\left(1-\frac{2}{5}\nu^2\right)e^{2\nu\bar\phi} \right]=20k^2+\frac{40k}{\nu^2(y_s-y)}+\frac{4(5-2\nu^2)}{\nu^4(y_s-y)^2} \,.
\end{equation}

The solution in the presence of a BH also includes the blackening factor
\begin{equation}
h(y) = 1  - \frac{\int_{-\infty}^y d\bar y \, e^{4 A(\bar y)}  }{ \int_{-\infty}^{y_h} d\bar y \, e^{4 A(\bar y)} }\,. \label{eq:sol_AdSnu_T}
\end{equation}
Then, the Hawking temperature and the entropy density of the BH are given by
\begin{eqnarray}
T_h &=& \frac{1}{4\pi} e^{-A(y_h)} |h^\prime(y)|_{y = y_h} =  \frac{k}{\pi}e^{-ky_h} (4 k y_s)^{-1/\nu^2} \frac{\lambda^{-3/\nu^2} e^{-\lambda}}{\Gamma\left[1-\frac{4}{\nu^2},\lambda\right]}   \,, \\
  s_h &=& \frac{4\pi}{\kappa^2} e^{-3 A(y_h)}  = \frac{4\pi}{\kappa^2} e^{-3 k y_h} (4 k y_s)^{-3/\nu^2} \lambda^{3/\nu^2}  \,,
\end{eqnarray}
%
%
%\begin{equation}gin{eqnarray}
%T_h &=& \frac{1}{4\pi} e^{-A(y_h)} |h^\prime(y)|_{y = y_h} =  \frac{k}{\pi}e^{-ky_h}e^{-4k(y_s-y_h)}\frac{[4k(y_s-y_h)]^{-4/\nu^2}}{\Gamma\left[1-\frac{4}{\nu^2},4k(y_s-y_h)\right]} \\
  %s &=& \frac{4\pi}{\kappa^2} e^{-3 A(y_h)}  = \frac{4\pi}{\kappa^2} e^{-3 k y_h} \left( 1 - \frac{y_h}{y_s} \right)^{\frac{3}{\nu^2}}  \,,
%\end{eqnarray}
%
%
where $\lambda\equiv 4k(y_s-y_h)$. This model, with $\nu = 1$, has been applied to particle physics in Refs.~\cite{Megias:2019vdb,Megias:2021arn} and to cosmology, including its thermodynamics and braneworld cosmology properties in Refs.~\cite{Fichet:2022ixi,Fichet:2022xol}. A study of these and/or other applications for other values of the parameter $\nu$ is left for future work.

%%%%%%%%%%%%%%%%%%%%%%%%%%%%%%%%%%%%%%%%%%
%% Optional
%%%%%%%%%%%%%%%%%%%%%%%%%%%%%%%%%%%%%%%%%%

\isPreprints{}{% This command is only used for ``preprints''.
\begin{adjustwidth}{-\extralength}{0cm}
} % If the paper is ``preprints'', please uncomment this parenthesis.

\printendnotes[custom] % Un-comment to print a list of endnotes

\reftitle{References}

\PublishersNote{}

\isPreprints{}{% This command is only used for ``preprints''.
\end{adjustwidth}
} % If the paper is ``preprints'', please uncomment this parenthesis.

\end{document}